\documentclass[aps,showpacs,twocolumn,superscriptaddress]{revtex4-1}

\usepackage{amsmath,amsfonts}
\usepackage{bbm}
\usepackage{graphicx}
\usepackage{color}
\usepackage{enumerate}
\usepackage{epsfig}
\usepackage{epstopdf}
\usepackage[caption=false]{subfig}

\newcommand{\e}{\mathrm{e}}
\newcommand{\up}{\uparrow}
\newcommand{\dn}{\downarrow}

\newcommand{\ket}[1]{\left| #1 \right\rangle}
\newcommand{\bra}[1]{\left\langle #1 \right|}
\newcommand{\braket}[2]{\left\langle #1 \right. \! \left| #2 \right\rangle}
\newcommand{\bracket}[3]{\left\langle #1 \right| #2 \left| #3 \right\rangle}
\newcommand{\vect}[1]{\boldsymbol{#1}}

\newcommand{\invisible}[1]{}

\allowdisplaybreaks

\newcommand{\eqnlab}[1]{\label{eqn:#1}}
\newcommand{\figlab}[1]{\label{fig:#1}}

\newcommand{\eqnref}[1]{(\ref{eqn:#1})}
\newcommand{\figref}[1]{\ref{fig:#1}}

\newcommand{\Eqnref} [1]{Eq.\ (\ref{eqn:#1})}

\newcommand{\beq}{\begin{equation}}
\newcommand{\beqn}{\begin{equation*}}
\newcommand{\eeq}{\end{equation}} 
\newcommand{\eeqn}{\end{equation*}}
\newcommand{\beqa}{\begin{eqnarray}}
\newcommand{\beqan}{\begin{eqnarray*}}
\newcommand{\eeqa}{\end{eqnarray}}
\newcommand{\eeqan}{\end{eqnarray*}}
\newcommand{\bdm}{\begin{displaymath}}
\newcommand{\edm}{\end{displaymath}}

\date{\today}

\begin{document}


\title{Controllable spin entanglement production in a quantum spin Hall ring}

\author{Anders Str\"{o}m}
\affiliation{Institute for Mathematical Physics, TU Braunschweig, D-38106 Braunschweig, Germany}
\author{Henrik Johannesson}
\affiliation{Department of Physics, University of Gothenburg, SE-412 96 Gothenburg, Sweden}
\author{Patrik Recher}
\affiliation{Institute for Mathematical Physics, TU Braunschweig, D-38106 Braunschweig, Germany}
\affiliation{Laboratory for Emerging Nanometrology Braunschweig, D-38106 Braunschweig, Germany}
% ----------------------------------------------------------------------------

\begin{abstract}
We study the entanglement production in a quantum spin Hall ring geometry where electrons of opposite
spins are emitted in pairs from a source and collected in two different detectors. Postselection of coincidence detector events gives rise to entanglement in the system, measurable through correlations between the outcomes in the detectors. We have chosen a geometry such that the entanglement depends on the dynamical phases picked up by the edge states as they move around the ring. In turn, the dependence of the phases on gate potential and Rashba interaction allows for a precise electrical control of the entanglement production in the ring.
\end{abstract}

% ----------------------------------------------------------------------------

\maketitle

% ----------------------------------------------------------------------------
\section{Introduction}

Quantum information processing relies on entanglement as its basic resource, with a high demand for viable and efficient schemes for producing and detecting quantum entangled states \cite{ScienceReview, Nielsenbook}. Much of the original research has focused on how to create two-electron entanglement in the solid state, with the aim to explore various paths towards scalable devices for quantum information processing. In equilibrium, spin entanglement can be produced controllably using pairs of quantum dots in the Coulomb blockade regime coupled to each other \cite{Loss1998, Burkard1999} or to a common superconductor \cite{Choi2000}.
However, in non equilibrium situations, pairwise spin-entangled electrons could be transported and the spin entanglement could be measured using similar means as in quantum optics with an appropriate spin to charge conversion \cite{Burkard2000, Kawabata2001, Chtchelkatchev2002}. Such ``entanglers" have been proposed in superconductor-normal junctions \cite{Recher2001, Lesovik2001, Recher2002, Bena2002, Recher2003, LevyYeyati2007, Cayssol2008, Sato2010} enjoying recent experimental support \cite{Schoenenberger2009, Strunk2010, Chandrasekhar2010, Heiblum2012} or in a three-terminal quantum dot device \cite{Yamaguchi2002}. Other schemes use a laser field \cite{Imamoglu}, or Kondo scattering by a magnetic impurity \cite{Costa,Sodano}. Besides spin, entanglement in the orbital sector of Cooper pairs \cite{Samuelsson1}, in Hanbury Brown and Twiss charge interferometers \cite{Samuelsson2} or carried by electron-hole pairs \cite{Beenakker} created by a tunnel junction, has been proposed. 

To guide electrons in solid-state systems, one-dimensional channels in integer quantum Hall devices are promising candidates and together with a controlled particle injection, tunneling junctions taking the role of linear optics beam splitters, and detection via correlation measurements, an all electronic table-top analog of photonics has been realized \cite{Martin}. 

More recently, there have been suggestions that the helical edge states in a two-dimensional quantum spin Hall insulator can also be used as electronic wave guides. These states, coming in Kramers' pairs of counterpropagating electrons with opposite spins, are topologically protected from elastic backscattering in the absence of time-reversal symmetry breaking \cite{HasanKane,QiZhang}. Their intrinsic helicity are readily usable for detecting spin entanglement \cite{Sato2010, Chen2012, Sato20141} or even for creating entanglement $-$ employing an Aharonov-Bohm flux to entangle the electrons (so-called time-bin entanglement) when injected into the quantum spin Hall insulator device \cite{Hofer}, or applying gate electrodes as beam splitters \cite{inhoferbercioux, Sato20142}. The latter two proposals have their analog in devices with quantum Hall edge states \cite{Beenakker, Samuelsson2}.

 \begin{figure}
	\begin{center}
		\includegraphics[width=0.7 \columnwidth]{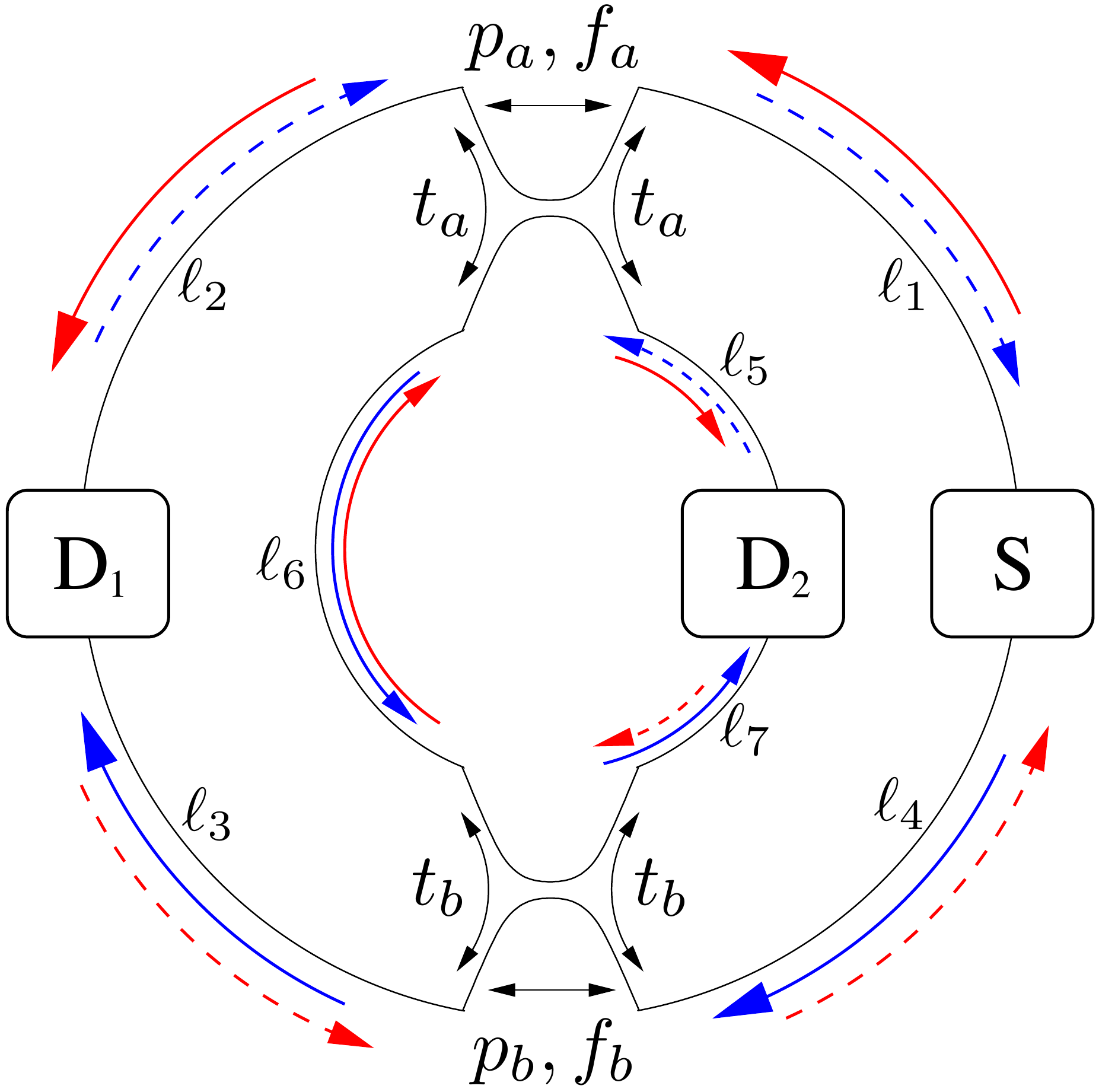}
		\caption{(Color online) Sketch of the setup, showing the source S and the two detectors D$_1$ and D$_2$ connected to a QSH ring. The two tunneling junctions have amplitudes $p_i$, $f_i$ and $t_i$ ($i\in\{1,2\}$) for spin-preserving tunneling, spin-flip tunneling and transmission, respectively. The length of the different parts of the edges are denoted $l_1$ through $l_7$. Red and blue arrows denote the travel direction of spin-up and spin-down electrons, respectively. Solid arrows are used for paths where the electrons originate from the source and end up in the detectors.}
		\figlab{corbino}
	\end{center} 
\end{figure}

In this paper, we suggest a setup for producing entangled pairs of electron spins in a quantum spin Hall ring, using gate electrodes to control $-$ with high precision $-$ the amount of entanglement produced in each pair. Any source producing pairs of electrons with opposite spins can be used with our design, provided that each electron comes with a definite spin. The electrons thus injected into the device are completely unentangled, with the entanglement of the outgoing electron states instead originating from a proper postselection \cite{Lebowitz, Bose02, lebedev} of detection events. 

Our setup provides an easily accessible means to control the entanglement production by shifting the phases of the plane waves representing the electrons moving around the ring. This can be performed by tuning the voltage of a backgate, leading to a shift of the phases via the resulting change of effective gate potentials felt by the electrons and strength of the Rashba spin-orbit interaction intrinsic to a two-dimensional electron system confined in a quantum well \cite{Winkler, Review}. Furthermore, we show that the possibility to individually tune the spin-flipping and spin-preserving tunneling amplitudes of the tunneling junctions of the setup allows for a full control of the quantization axes of the entangled spin states: any linear combination of the four possible Bell states are achievable as output states of the device. This may pave the way for experimental tests of Bell inequalities using electron spins, and also has potential as a resource for quantum information purposes.
 
In the next section, Sec.\ II, we present the design of the setup, write down the associated Hamiltonian and introduce a scattering matrix formalism, enabling us to monitor how the outgoing electron states change as the effective gate potential, Rashba interaction, and tunneling amplitudes are varied. The scattering matrix approach is put to use in Sec.\ III, where we carry out a detailed analysis of the amount of entanglement in the outgoing states. In Sec.\ IV, we address the issue how to experimentally measure  the entanglement produced by our device. Sec.\ V, finally, contains a brief summary.

\section{Model}
\subsection{Setup}
We consider a setup consisting of a ring formed in a HgTe quantum well supporting a quantum spin Hall (QSH) state \cite{HasanKane,QiZhang}. The ring, sketched in Fig. \figref{corbino}, is separated into two halves joined together by two tunneling junctions. Moreover, as also shown in Fig. 1, a source (S) and two detectors (D$_1$ and D$_2$) are connected to the edges of the ring. A closed ring in topological insulators based on HgTe quantum wells has previously been investigated in Ref. \cite{patrikpaolo}, in a different context. 
In a QSH system, the edge states are helical, meaning that counterpropagating electrons are Kramers partners, related by time reversal symmetry (TRS). 
In the realization of a QSH phase in a HgTe quantum well, the spinor components of the edge states are labeled by a quantum number taking two values, call them $\pm$, depending on which linear combination of total angular momentum states they are built from \cite{MaciejkoReview}. By choosing the spin quantization axis along $\bra{+}  \vect{S}  \ket{+}$, with  $\vect{S}$ the spin operator of an electron, the orthogonal $\ket{+}$ and $\ket{-}$ states will correspond to spin-up and spin-down eigenstates (related by time reversal), and can be relabeled $\ket{\up}$ and $\ket{\dn}$, respectively. In other words, particles emitted from the source S with spin along $\bra{+}  \vect{S}  \ket{+}$ will enter only $\ket{\up}$ states, conversely the ones with opposite spin will enter only $\ket{\dn}$ states. It is important to note, however, that $\bra{+}  \vect{S}  \ket{+}$ depends on the energy of an electron \cite{patrikpaolo, Virtanen12, Schmidt12}, and therefore, using it to define a spin quantization direction is meaningful only in a narrow energy range. To have a common spin-quantization axis for the electrons thus requires that they all leave the source S with roughly the same energy.

Sources for single spin-polarized electron pairs emitted into helical edge states have been recently proposed in Refs.\  \cite{Hofer} and \cite{inhoferbercioux}, where a periodically driven quantum dot connected to a helical edge emits a pair of electrons every time the highest occupied energy level of the dot crosses the Fermi level of the edge from below.
We propose a similar source, but with two separate quantum dots capacitively connected to a metallic gate with their energy levels aligned, emitting one electron each into the helical edge via a tunneling junction. Having double single-particle sources \cite{Splettstoesser, Bocquillon} ensures that the emitted electrons are unentangled, a crucial feature of our setup. As depicted in Fig.\ \figref{source}, a drain is placed between the two dots to ensure that only electron pairs with opposite spins are injected into the ring. The proposed source is sketched in Fig.\ \figref{source} and corresponds to the source S in Fig.\ \figref{corbino}. 

 \begin{figure}
	\begin{center}
		\includegraphics[width=0.5 \columnwidth]{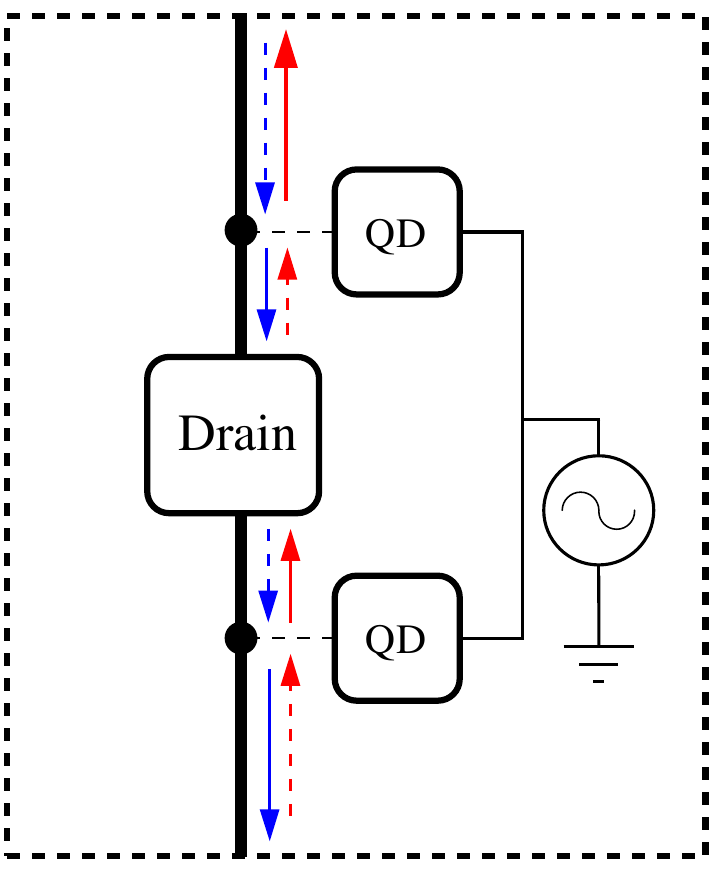}
		\caption{(Color online) Sketch of the source S (cf.\ Fig.\ 1). The two quantum dots (QDs) emit single electrons simultaneously into the helical edge when their energy levels cross the Fermi level of the edge from below. If the upper QD releases a spin-up electron, and the lower QD a spin-down electron, the device produces a pair of opposite spins traveling away from the source. Otherwise, an electron with unwanted spin orientation is collected in the drain.}
		\figlab{source}
	\end{center} 
\end{figure}

In Fig.\ \figref{corbino}, electrons emitted from the source S are assumed to move counterclockwise along the outer edge of the ring if their spins are up, and clockwise if their spins are down (and vice versa on the inside of the ring). Assuming now that all electrons emitted from S have the same energy, there are then two different emitted single-electron states, which we denote $a^\dag_{S\up}\ket{0}$ and $a^\dag_{S\dn}\ket{0}$, with $\ket{0}$ the filled Fermi sea of edge electrons defining the ground state. An electron exiting the ring can have either spin up or spin down as it enters either detector D$_1$ or detector D$_2$. There are thus four possible outgoing states in the detectors, $b^\dag_{j,\sigma}\ket{0}$, where $j=1,2$ and $\sigma=\up,\dn$.

At junctions $a$ and $b$  (the upper and lower junction in Fig.\ \figref{corbino}, respectively), electrons can either tunnel between the left and right halves of the ring through the junction, or be transmitted past the junction and stay on the same half. In the tunneling junctions, the two edges are squeezed together, allowing for scattering between the edges \cite{krueckl, dolcini}. Without breaking TRS, the electrons can either stay on the same edge, keeping its spin, or scatter to the opposite edge where states of both spins (and therefore both directions) are available. In the general case, there is thus a finite amplitude $p_i$ for spin-preserving tunneling and a finite amplitude $f_i$ for spin-flip tunneling ($i=a,b$). The amplitude for transmission past the junction, between the inner and outer side of the ring, is denoted $t_i$. In this process the spin is conserved, assuming the quantization axes on the inner and outer sides of the ring to be the same. Note that depending on the exact configurations of the different potential energies experienced by the electrons in the various scattering processes, the notions of ``tunneling" and ``transmission" may be interchanged. For simplicity, we will keep the terms as introduced above for all possible configurations of potentials. The probabilities for the possible processes of an electron arriving at one of the junctions must add up to one,
\beq
|t_a|^2+|p_a|^2+|f_a|^2=|t_b|^2+|p_b|^2+|f_b|^2=1.\eqnlab{one}
\eeq
\noindent  By symmetry, with no bias applied, $|p_i|$ and $|f_i|$ are the same regardless of whether electrons tunnel from left to right or vice versa. For the two different (left and right) transmission amplitudes at each junction, \Eqnref{one} therefore implies that $ |t_{\mathit i, \mathrm{left}}|^2=|t_{\mathit i, \mathrm{right}}|^2$. We assume furthermore that the extra phase contribution from transmission past the junction is negligible, so that $t_{\mathit i,\mathrm{left}}=t_{\mathit i, \mathrm{right}}=t_i$ holds.

\subsection{Hamiltonian}

To analyze the physics of the setup, we shall use time-independent scattering theory with the energy fixed to the Fermi level of the edge states. Setting the stage, we introduce  coordinates $x_1, x_2,..., x_7$ for the different edge segments of length $l_1,l_2,..., l_7$, respectively (see Fig. 1). Imposing open boundary conditions for each segment, we write the corresponding single-particle Hamiltonians (with $\hbar=1$) as \begin{equation}
H_j = -iv_F\partial_{x_j}\sigma_z -i\alpha \partial_{x_j} \sigma_y - eV_g, \ \ j = 1, 2, ..., 7,
\end{equation}
having linearized the edge state dispersion about the Fermi points $\pm k_F$. 
\noindent Here $v_F$ is the Fermi velocity, and $\sigma_y$ and $\sigma_z$ are Pauli matrices acting on the spin states $\ket{\up} = (1\, 0)^{T}$ and $\ket{\dn} = (0\, 1)^{T}$. The second term of $H_j$ encodes the Rashba spin-orbit interaction of strength $\alpha$, with the third term a potential term, where $V_g$ denotes the effective gate potential 
felt by the electrons.

The Rashba coupling $\alpha$ has a complex dependence on
several distinct features of the quantum well in which the ring is defined \cite{MGJJ}. Like the effective gate potential, it depends in particular on the electric field from any top- or bottom gates applied to the device, and can thus be tuned by tuning the gate potentials. It is convenient to absorb the Rashba interaction into the kinetic energy via the unitary transformation $U=\exp(-i\sigma_x \theta/2)$, and write the Hamiltonian as $H_{j}'=UH_{j}U^{\dagger}$ with
 \beq \label{rotatedH}
 H_j'= -iv_\alpha \partial_{x_j} \sigma_{z'} -  eV_g, \ \ j = 1,2,...,7
 \eeq
 \noindent with $v_\alpha=\sqrt{v_F^2+\alpha^2}$, and with the spin-quantization axis $\hat{z}'$ inclined by an angle $\theta = \arcsin(\alpha/v_F)$ with respect to the $\hat{z}$-axis which defined the original spin-quantization axis.  In this way, we are dealing with pure helical edge states with a renormalized Fermi velocity $v_{\alpha}$ and a new spin-quantization axis $\hat{z}'$.

Given a wave number $k$, the corresponding helical eigenstate of $H_j'$, 
\beq  \label{up}
\psi_{k,\up}(x_j) = \e^{ikx_j} \ket{\up},
\eeq
\noindent has a Kramers' partner
\beq  \label{down}
\psi_{-k,\dn}(x_j) = \e^{-ikx_j} \ket{\dn}, 
\eeq
both with energy
\beq  \label{energy}
E = k v_{\alpha} - eV_g.
\eeq

Joining together the open boundaries of the segments at the junctions $a$ and $b$ so as to form the ring in Fig.\ 1, electrons can tunnel or be transmitted from one segment to another. In the present case of elastic tunneling processes, an electron in a state with wave number $\pm k$ will then emerge on the other side of the junction in a state with the same or the opposite wave number, corresponding to spin-flip and spin-preserving tunneling, respectively. Likewise, a transmitted electron is simply transferred between states with the same wave number. Independent of the type of process or the direction of motion, it follows that the plane wave representing an electron having traveled a distance $l$ from the source S will be phase shifted by an amount $(E+eV_g)l/v_\alpha$, not counting the phase shifts acquired in the tunneling processes. Since the phase $(E+eV_g)l/v_\alpha$ is proportional to the distance traveled by the electron, the geometry of the ring becomes crucial. Importantly, the path $l_6$ is only traveled by electrons which have flipped their spin. As will be explained in the next section, this fact allows for controlling the entanglement in the postselected states that originate from a Kramers' pair emitted from the source S. 

Before proceeding,  let us pause and recall that the part of the dynamical phase that depends on the Rashba SO coupling is spin dependent in ordinary spinful electron liquids. This is somewhat related to the dynamical part of the Aharonov-Casher effect \cite{ac1}, where electron spins moving around a uniformly charged thread acquire a phase due to the electric field (dual to the Aharonov-Bohm effect where electron charges move around a magnetic flux and acquire a phase due to the magnetic field). In this situation, the electric field also affects the dynamical phases of the electrons, just as in the case of a Rashba SO coupling. This `dynamical part of the Aharonov-Casher effect" from the Rashba effect has been experimentally detected in a HgTe quantum well \cite{acexp}.

\subsection{$S$-matrix}
Following B\"uttiker \cite{buttiker2}, we introduce second-quantized operators $a^\dag_{j \sigma}$ and $a_{j \sigma}, $ $j=${\small S, D$_1$, D$_2$}; $\sigma = \up, \dn$, which  create and annihilate electrons in the ``incoming" states (i.e., the states emitted from the source and the detectors before hitting a junction). Similarly, $b^\dag_{j \sigma}$ and $b_{j \sigma}$, j=\! {\small S, D$_1$, D$_2$}; $\sigma = \up, \dn$, create and annihilate electrons in the ``outgoing" states (i.e. the scattered states leading to the detectors or the source).
The $a$ and $b$ operators are connected through the scattering relation
\beq  \label{Smatrix}
\begin{pmatrix} b_{1\up}\\ b_{1\dn}\\ b_{2\up}\\ b_{2\dn}\\b_{S\up}\\ b_{S\dn}\end{pmatrix}= S  \begin{pmatrix} a_{1\dn}\\ a_{1\up}\\ a_{2\dn}\\ a_{2\up}\\a_{S\dn}\\ a_{S\up} \end{pmatrix},
\eeq

\noindent with the scattering matrix $S$ providing a unitary mapping of the $a$ operators into the $b$ operators. The full $S$-matrix is presented in the Appendix [see \Eqnref{fullsmatrix}]. If the temperature is low enough we can neglect incoming electrons from the detectors, and we can focus on the part of the 
$S$-matrix which maps the $a$ operators for states emitted from the source into the $b$ operators for the scattered states. We call this the {\em reduced scattering matrix} $\tilde S$. From Eq. (A.2) in the Appendix we read off:
\beq
\begin{pmatrix} b_{1\up}\\ b_{1\dn}\\ b_{2\up}\\ b_{2\dn}\\b_{S\up}\\b_{S\dn}\end{pmatrix}=\tilde S \begin{pmatrix} a_{S\up} \\ a_{S\dn} \end{pmatrix},
\eeq

\noindent with
\beq  \label{tilde}
\tilde S = \begin{pmatrix} p_a e^{-iK(l_1+l_2)} &  f_bt^*_a e^{-iK(l_2+l_4+l_6)} \\ f^*_at_b e^{-iK(l_1+l_3+l_6)} & p^*_b e^{-iK(l_3+l_4)} \\ -t_a e^{-iK(l_1+l_5)} &  f_bp^*_a e^{-iK(l_4+l_5+l_6)} \\ -f^*_ap_be^{-iK(l_1+l_6+l_7)} &  t^*_b e^{-iK(l_4+l_7)}\\ -f_a^*f_b e^{-iK(l_1+l_4+l_6)} & 0\\ 0 & -f_a^*f_be^{-iK(l_1+l_4+l_6)}\end{pmatrix}.
\eeq

\noindent Here $K\equiv (E+eV_g)/v_\alpha$, with the relative minus signs and the complex conjugations of the tunneling and transmission amplitudes in Eq.\ (\ref{tilde}) prescribed by the unitarity of the full $S$-matrix in Eq.\ (\ref{Smatrix}). Note that the phase factor $\exp(iKl_6)$ occurs only in the matrix elements for spin-flipping processes.

\section{Entanglement}

\subsection{Postselection}
When the source emits two unentangled electrons of opposite spins, the incoming two-electron state is $\ket{\Psi_{in}} = a^\dag_{S \up} a^\dag_{S \dn} \ket{0}$, with $\ket{0}$ the filled Fermi sea of edge electrons. By inspection of the S-matrix, with four possible scattering channels for each electron, we can easily read off the outgoing two-electron state, calling it $\ket{\Psi'_{\textit out}}$. Using that $S^{-1}=S^\dag$, it follows from Eq.\ (\ref{tilde}) and \Eqnref{fullsmatrix} that
\beq
 \begin{pmatrix} a_{S\up} \\ a_{S_\dn} \end{pmatrix} = \tilde{S}^\dag \begin{pmatrix} b_{1\up}\\ b_{1\dn}\\ b_{2\up}\\ b_{2\dn}\\b_{S\up}\\b_{S\dn}\end{pmatrix},
 \eeq
 
 \noindent and we obtain
\begin{multline}  \label{psioutnprime}
\ket{\Psi'_{\textit out}}\\
=N'\Big(p_a e^{-iK(l_1+l_2)}b^\dag_{1\up}+f_a^*t_be^{-iK(l_1+l_3+l_6)}b^\dag_{1\dn}\\
-t_a e^{-iK(l_1+l_5)}b^\dag_{2\up}- f_a^*p_b e^{-iK(l_1+l_6+l_7)}b^\dag_{2\dn}\\
- f_a^*f_b e^{-iK(l_1+l_4+l_6)}b^\dag_{S\up}\Big)\\
\times\Big( f_b t_a^* e^{-iK(l_2+l_4+l_6)}b^\dag_{1\up}+  p_b^*e^{-iK(l_3+l_4)}b^\dag_{1\dn}\\
+f_bp_a^*e^{-iK(l_4+l_5+l_6)}b^\dag_{2\up}+t_b^* e^{-iK(l_4+l_7)}b^\dag_{2\dn}\\
- f_a^*f_b e^{-iK(l_1+l_4+l_6)}b^\dag_{S\dn}\Big)\ket{0},
\end{multline}

\noindent where $N'$ is a normalization factor, determined by choosing $\braket{\Psi'_{\textit out}}{\Psi'_{\textit out}}=1$.

We project this state to the realization that each detector receives exactly one electron. This process is referred to as postselection \cite{Bose02, lebedev}. Thus, we keep only the terms in Eq. (\ref{psioutnprime}) where one particle gets detected in D$_1$ and the other in D$_2$. By using the antisymmetry of the $b_{j \sigma}^{\dagger}$ operators, we can express the resulting state on the form:
\begin{multline}  \label{psiout}
\ket{\Psi_{\textit out}}=N\Big( f_b\left(\left|p_a\right|^2+\left|t_a\right|^2\right) e^{-iK(l_{\up\up}+l_6)}\ket{\up\up}\\ 
+f_a^*\left(\left|p_b\right|^2+\left|t_b\right|^2\right)e^{-iK(l_{\dn\dn}+l_6)}\ket{\dn\dn}\\
+\left[ p_at_b^* e^{-iKl_{\up\dn}}+f_a^*f_bp_bt_a^* e^{-iK(l_{\up\dn}+2l_6)}\right]\ket{\up\dn}\\\
+\left[ p_b^*t_a e^{-iKl_{\dn\up}}+f_a^*f_bp_a^*t_b e^{-iK(l_{\dn\up}+2l_6)}\right]\ket{\dn\up} \Big),
\end{multline}

\noindent where the states $\ket{\sigma\sigma'}$ are defined as $b^\dag_{1\sigma}b^\dag_{2\sigma'}\ket{0}$, the lengths are defined through
\begin{align}
 l_{\up\up}&=l_1+l_2+l_4+l_5, \eqnlab{lupup}\\
l_{\dn\dn}&=l_1+l_3+l_4+l_7, \eqnlab{ldndn}\\
l_{\up\dn}&=l_1+l_2+l_4+l_7, \eqnlab{lupdn}\\
l_{\dn\up}&=l_1+l_3+l_4+l_5, \eqnlab{ldnup}
\end{align}

\noindent and the normalization factor is
\begin{multline}
N=\Big(\left|f_b(\left|p_a\right|^2+\left|t_a\right|^2)\right|^2+\left|p_at_b^*+f_a^*f_bp_bt_a^*e^{-2iKl_6}\right|^2\\
+\left|f_a^*(\left|p_b\right|^2+\left|t_b\right|^2)\right|^2+\left|p_b^*t_a+f_a^*f_bp_a^*t_be^{-2iKl_6}\right|^2\Big)^{-1/2},
\end{multline}

\noindent chosen so that $\braket{\Psi_\textit{out}}{\Psi_\textit{out}}=1$. This incorporates the postselection condition. Clearly, the different tunneling and transition amplitudes play an important role for the spin entanglement of $\ket{\Psi_{\textit out}}$ that we will study in the following. In the two cases where $(1)$ $f_a=f_b=0$ and $(2)$ $p_a=p_b=0$, the corresponding outgoing states are 
\begin{align}
\ket{\Psi_1}&=N\left( p_at_b^*e^{-iKl_{\up\dn}}\ket{\up\dn}+p_b^*t_ae^{-iKl_{\dn\up}}\ket{\dn\up}\right)\eqnlab{psi1},\\
\begin{split}\ket{\Psi_2}&=N( f_b\left|t_a\right|^2e^{-iK(l_{\up\up}+l_6)}\ket{\up\up}\\
&\phantom{=}\ +f_a^*\left|t_b\right|^2e^{-iK(l_{\dn\dn}+l_6)}\ket{\dn\dn}),\end{split}\eqnlab{psi2}
\end{align}

\noindent respectively, which are maximally entangled Bell states if the two junctions are equal (i.e.,\ $|f_a|=|f_b|$, $|p_a|=|p_b|$, and $|t_a|=|t_b|$). Setting $|t_a|=|t_b|=0$ also produces the state $\ket{\Psi_2}$, with  $t_i$ in \Eqnref{psi2} replaced by $p_i$. This means that if we are able to separately tune the amplitudes for spin-flipping and spin-preserving tunneling through the junctions, we can choose the states going out of the device freely, from the opposite-spin Bell state $\Psi_1$ to the same-spin Bell state $\Psi_2$, as well as any superposition of the two. Although not yet experimentally verified, such control of the different amplitudes in tunneling junctions of helical edges -- using local electrical gates to control the amount of spin-orbit coupling experienced by the electrons in the junctions -- has previously been proposed in Refs.\ \cite{krueckl, dolcini}.

Importantly, by tuning the effective gate potential $eV_g$ and/or the Rashba coupling $\alpha$ to control the relative phases between the $\ket{\up\dn}$ and $\ket{\dn\up}$ terms in \Eqnref{psi1} and the $\ket{\up\up}$ and $\ket{\dn\dn}$ terms in \Eqnref{psi2}, allows for choosing the outgoing states to be any linear combination of {\em all four} of the standard Bell states, $\ket{\Psi_1^\pm}=(\ket{\up\dn}\pm\ket{\dn\up})/\sqrt{2}$ and $\ket{\Psi_2^\pm}=(\ket{\up\up}\pm\ket{\dn\dn})/\sqrt{2}$. This is a particular advantage of our device, made possible by its special geometry. Below we shall explore in detail how it can be used to yield a precise control of the spin entanglement production. We note that tunable spin-entanglement production has also been proposed in a \emph{non-helical} Mach-Zehnder setup with Rashba spin-orbit interaction leading to spin-rotation \cite{Zuelicke2005} different to our proposal.

\subsection{Efficiency}

   \begin{figure}[ht!]
   \begin{center}
   \subfloat[]{
   \includegraphics[width=0.9\columnwidth]{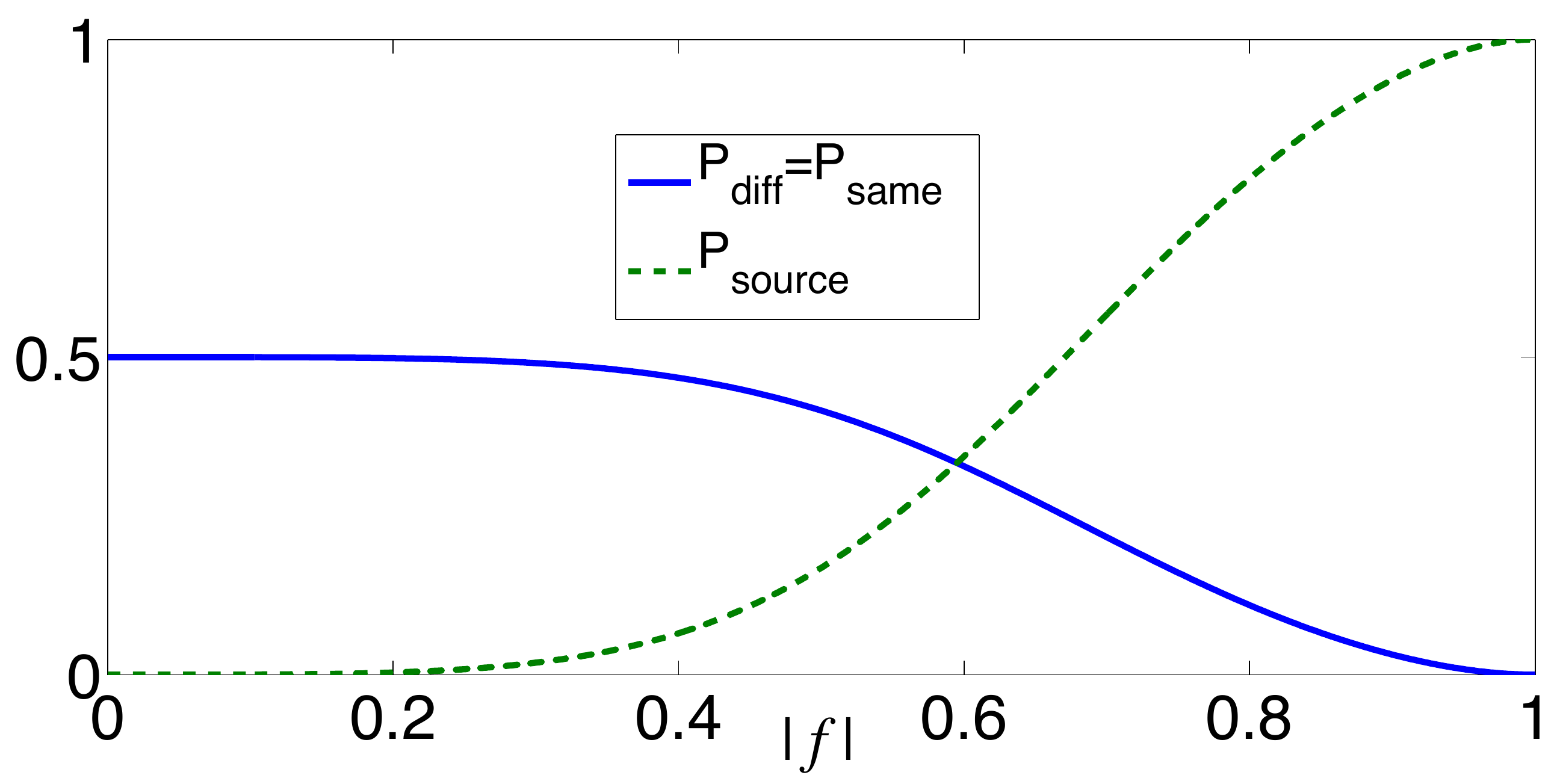}\figlab{visibility_a}
   }\\
   \subfloat[]{
     \includegraphics[width=0.9\columnwidth]{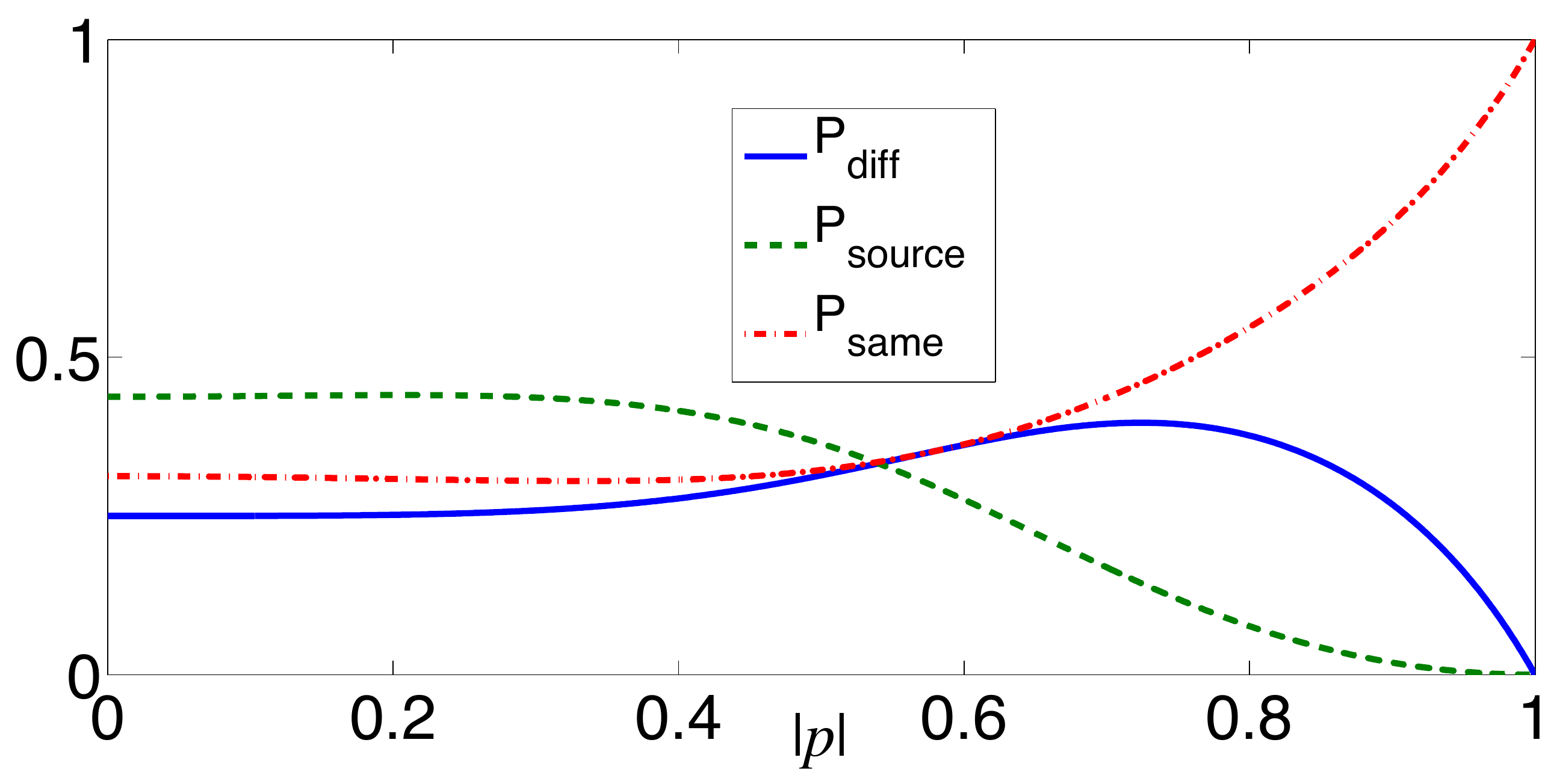}\figlab{visibility_b}
     }
\caption{(Color online) Probabilities (i) for the electrons to end up in different detectors $(P_{\textrm{diff}})$, (ii) for at least one electron returning back to the source $(P_{\textrm{source}})$, and (iii) for both electrons being collected in the same detector $(P_{\textrm{same}})$ as a function of (a) the spin-flip and (b) the spin preserving tunneling amplitudes. In (a) $p=t$, and in (b) $f=t$. Maximum entanglement is always produced when $f=0$ or $p=0$, and both cases are supported by a finite $P_{\textrm{diff}}$ (blue curve).}
   \figlab{visibility}
   \end{center}
   \end{figure}

One must note that the postselection of only certain events will decrease the efficiency of the setup, and it is essential to calculate how large the fraction of postselected states is compared to the discarded ones.  Let us call the probability for ending up with at least one electron in the source $P_{\textrm{source}}$, the probability for detecting one electron in each detector $P_{\textrm{diff}}$, and have $P_{\textrm{same}}$ denoting the probability for ending up with both electrons in the same detector. No other options are available, so $P_{\textrm{source}}+P_{\textrm{diff}}+P_{\textrm{same}}=1$. In the case of equal junctions, i.e., $f_a=f_b\equiv f$, $p_a=p_b\equiv p$, and $t_a=t_b\equiv t$, Eq.\ (\ref{psioutnprime}) implies that the probability for the two electrons to end up in different detectors is
\begin{equation}
P_{\textrm{diff}}=(N')^2\left[2PT\left| 1\!-\!F e^{-2iKl_6}\right|^2\!+\!2F(P\!-\!T)^2\right],\eqnlab{pdiff}
\end{equation}

\noindent where the tunneling and transmission probabilities $F\equiv |f|^2$, $P\equiv |p|^2$ and $T\equiv |t|^2$ have been introduced for convenience, while
\begin{multline}
P_{\textrm{source}}=(N')^2\left[F^4+2F^2(FP + P + FT + T)\right],
\end{multline}

\noindent and
\begin{multline}
P_{\textrm{same}}=(N')^2\Big( \left| P\!+\!FTe^{-2iKl_6}\right|^2\!+\!\left| T\!+\!FPe^{-2iKl_6}\right|^2\Big).\eqnlab{psame}
\end{multline}

\noindent An expression for $(N')^2$ can now be read off from Eqs.\ \eqnref{pdiff}-\eqnref{psame}, using the fact that the probabilities sum up to one. Varying $f$ between $0$ and $1$, while keeping $p=t$, gives $P_{\textrm{diff}}=0.5$ when $f=0$. The choice $f=1$ causes all electrons to travel back to the source, rendering $P_{\textrm{diff}}=0$. Varying $p$ ($t$) between $0$ and $1$, while keeping $f=t$ ($f=p$), gives $P_{\textrm{diff}}=0.25$ when $p=0$ ($t=0$), increasing to its maximum value $P_{\textrm{diff}}=0.40$ when $p=0.75$ ($t=0.75$). When $p$ ($t$) grows larger, the probability to end up in D$_1$ (D$_2$) will approach unity, which means that $P_{\textrm{diff}}=0$ also for $p=1$ ($t=1$). This is illustrated in Fig.\ \figref{visibility}. 

If a source with two quantum dots like the one in Fig.\ \figref{source} is used, with no preferred spin orientation in the dots, only $1/4$ of the produced pairs will be useful for the entangler. In this case, $P_{\textrm{diff}}$ is further reduced by a factor of $4$. As an example, in the case of small $p$, around $6$\% of the produced electrons are left available for entanglement. We can compare this to the setup of Ref.\ \cite{inhoferbercioux}, where the number of entangled electrons instead approach zero in the limit of maximal spin entanglement.

\subsection{Concurrence}
To estimate the entanglement produced by the post-selection of the detected states, we use the concurrence $C$ as an entanglement measure \cite{wootters}. For a pure state of a bipartite system $\ket{\Psi}$, this is defined as
\beq
C=\left| \bracket{\Psi}{\sigma_y\otimes\sigma_y}{\Psi^*}\right|,\eqnlab{conc}
\eeq

\noindent where the complex conjugate is to be taken in the basis in which the Pauli matrices are written. We let $A_{\sigma\sigma'}$ denote the amplitude for the $\ket{\sigma\sigma'}$ state, so that 
\beq
\ket{\Psi_{\textit out}}=N\left(A_{\up\up}\ket{\up\up} +A_{\up\dn}\ket{\up\dn}+ A_{\dn\up}\ket{\dn\up}+ A_{\dn\dn}\ket{\dn\dn}\right)\eqnlab{app}\eeq

\noindent and
\beq
C=2N^2\left| A_{\up\dn}A_{\dn\up}-A_{\up\up}A_{\dn\dn} \right|.\eqnlab{concurrence}
\eeq

\noindent As can be read off from Eqs. \eqnref{lupup}-\eqnref{ldnup}, $l_{\up\dn}+l_{\dn\up}=l_{\up\up}+l_{\dn\dn}$, and it then follows from Eq. (\ref{psiout}) that the concurrence can be written as
\begin{multline}  \eqnlab{cac}
C=2N^2\Big| \big( f_a^*f_b\left[|p_a|^2|t_b|^2+|p_b|^2|t_a|^2\right]e^{-2iKl_6}\\
-f_a^*f_b\left[|p_a|^2+|t_a|^2\right]\left[|p_b|^2+|t_b|^2\right]e^{-2iKl_6}\\
+p_ap_b^*t_at_b^*+f_a^{*2}f_b^2p_a^*p_bt_a^*t_be^{-4iKl_6}\big)e^{-iK\left(l_{\up\up}+l_{\dn\dn}\right)}\Big|\\
=2N^2\big| p_ap_b^* t_at_b^*+p^*_ap_bt_a^*t_bf_a^{*2}f_b^2e^{-4iKl_6}\\
-f_a^*f_b\left(|p_a|^2|p_b|^2+|t_a|^2|t_b|^2\right)e^{-2iKl_6}\big|
\end{multline}

\noindent and we see that the only length affecting the concurrence is $l_6$. 

Equation (26) shows that our setup allows for entanglement production of highly detailed control.
It is instructive to uncover the particular role of the phase factors $e^{\pm 2iKl_6}$ in \Eqnref{cac} for achieving this. First note that according to \Eqnref{concurrence}, the amount of entanglement is determined by the absolute value of the  difference between the products of the same-spin and opposite-spin state amplitudes. The phases of these products are proportional to the total distance traveled by the electrons. Our setup is designed so that the total length of the paths for two different states with either the same or opposite spins is always given by the length $l_6$ times the total number of spin-flips, plus $l_{\up\up}+l_{\dn\dn}$ (same-spin states) or $l_{\up\dn}+l_{\dn\up}$ (opposite-spin states). Now, Eqs.\  \eqnref{lupup} - \eqnref{ldnup} show that $l_{\up\up}+l_{\dn\dn}=l_{\up\dn}+l_{\dn\up}$, so that part of the phase factor will be common to the full expression within the absolute signs in \Eqnref{concurrence}. Since the electrons are injected into the edge with opposite spins, the same-spin and opposite-spin states will have experienced different numbers of spin flips. As a consequence, and as seen in \Eqnref{cac}), only phases proportional to $l_6$ will remain in the various terms in the expression for the concurrence $C$. These phases contain the parameter $K$, which makes them experimentally tunable via the gate voltage $V_g$ that parametrizes $K$: the effective gate potential $eV_g$ as well as the tunable part of the Rashba coupling $\alpha$ (which enters into the expression for $v_{\alpha}$) depend on $V_g$. We should here stress that it is the topology of the ring structure, where both junctions are directly connected to both detectors, that makes this phase dependence of the concurrence possible. In this context, we also point out that the expression for $C$ is gauge invariant, as it should be. This is most easily seen by going back to \Eqnref{cac}: the phase factor $\exp(-2iKl_6)$ appears always together with $f_a^* f_b$ which describes traversing the $l_6$-segment in two different directions. This phase is therefore gauge invariant \cite{footnote}. In the next section we shall elaborate on how the phase dependence of the concurrence can be exploited experimentally to control the quantum entanglement of the spins. 

Before doing so, however, let us study how the various tunneling amplitudes in  \Eqnref{cac} influence the concurrence. In Fig.\ \figref{conc_aniso}, the maximum concurrence obtainable by tuning the gate voltage $V_g$ is shown for a number of different configurations of the tunneling amplitudes. The $x$ and $y$ axes represent the ratio $r_f\equiv |f_a|/|f_b|$ and $r_p\equiv |p_a|/|p_b|$, respectively. Five different configurations of $f_a$, $p_a$, and $t_a$ are chosen for Figs.\ 4(a) - 4(e).  The important information to be read off from these figures is that for the case of equal junctions [represented by the $(r_f,r_p)=(1,1)$ corners of the figures], $V_g$ can always be chosen so as to obtain maximal entanglement ($C=1$). Moreover, most other configurations also give reasonably high values for $C$. The $(r_f,r_p)=(0,0)$ corners of the figures on the other hand represent the case where $|t_a|\rightarrow 1$, fixing the spins of the electrons ending up in D$_2$ and consequently making $C\rightarrow 0$. 
 
    \begin{figure}[ht!]
   \begin{center}
   \subfloat[$f_a=0.32$, $p_a=t_a=0.67$]{
   \includegraphics[width=0.49\columnwidth]{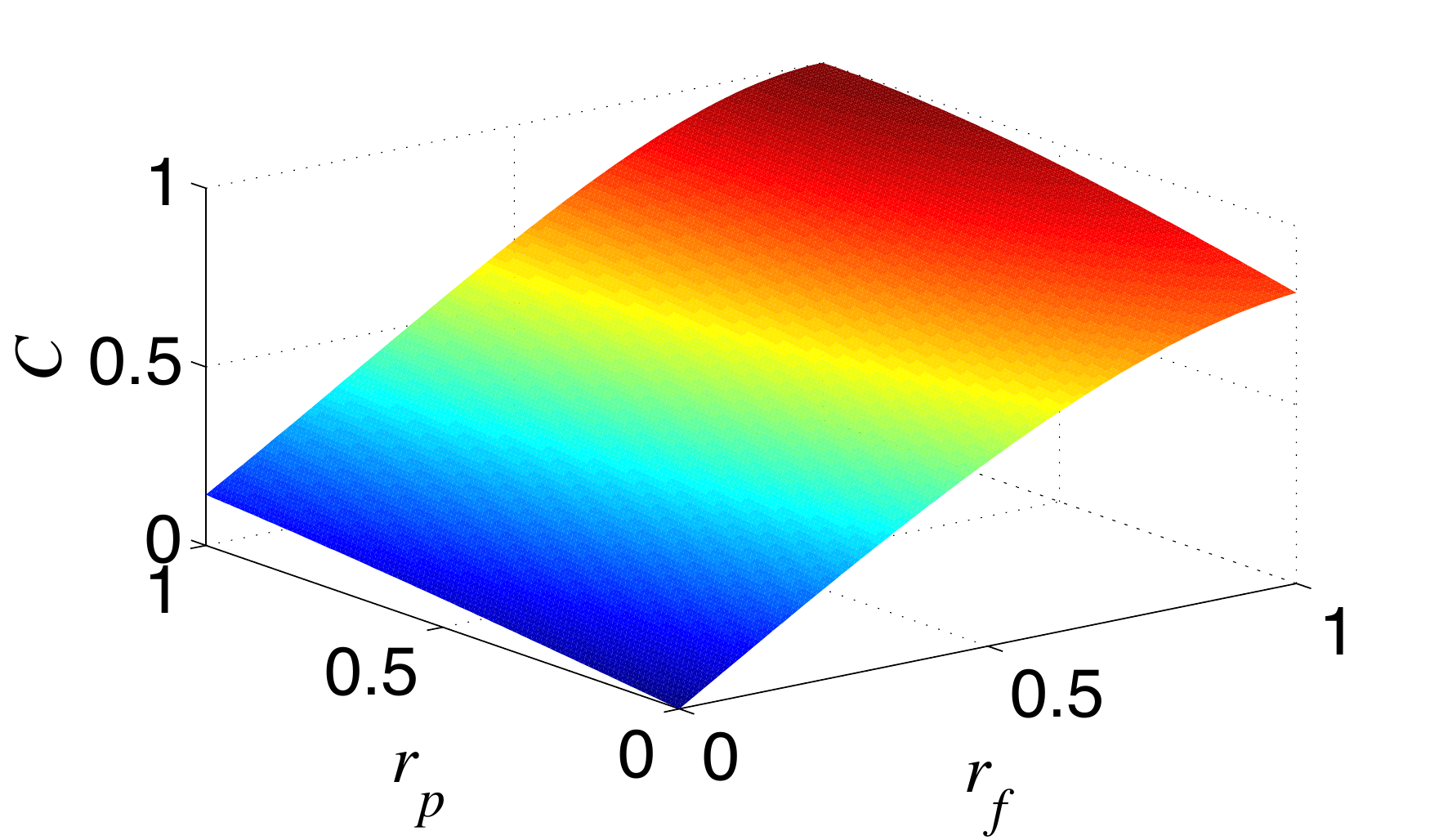}\figlab{conc_aniso_a}
   }
   \subfloat[$f_a=0.32$, $p_a=0.42$, $t_a=0.85$]{
     \includegraphics[width=0.49\columnwidth]{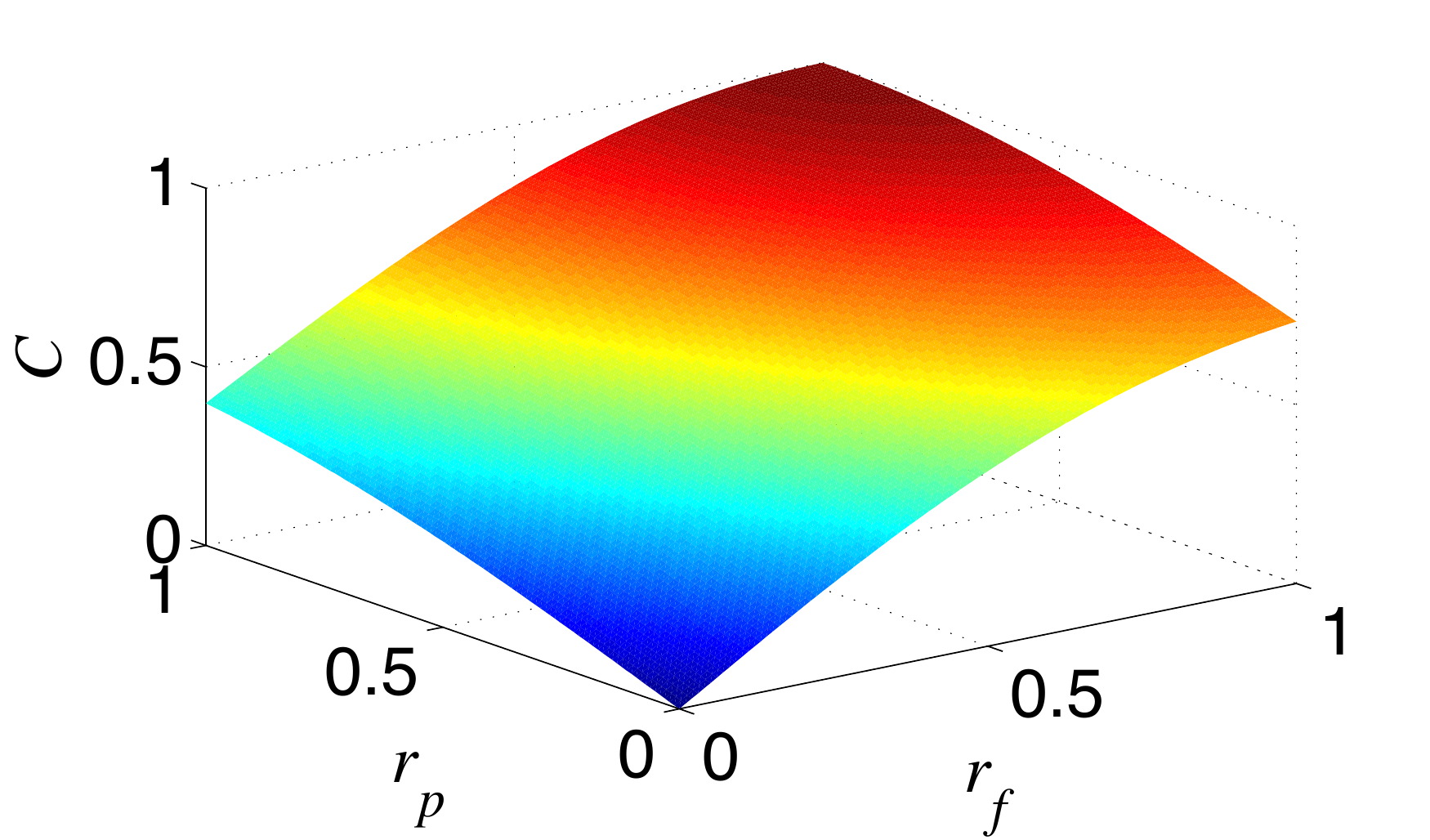}\figlab{conc_aniso_b}
     }\\
     \subfloat[$f_a=0.89$, $p_a=t_a=0.32$]{
        \includegraphics[width=0.49\columnwidth]{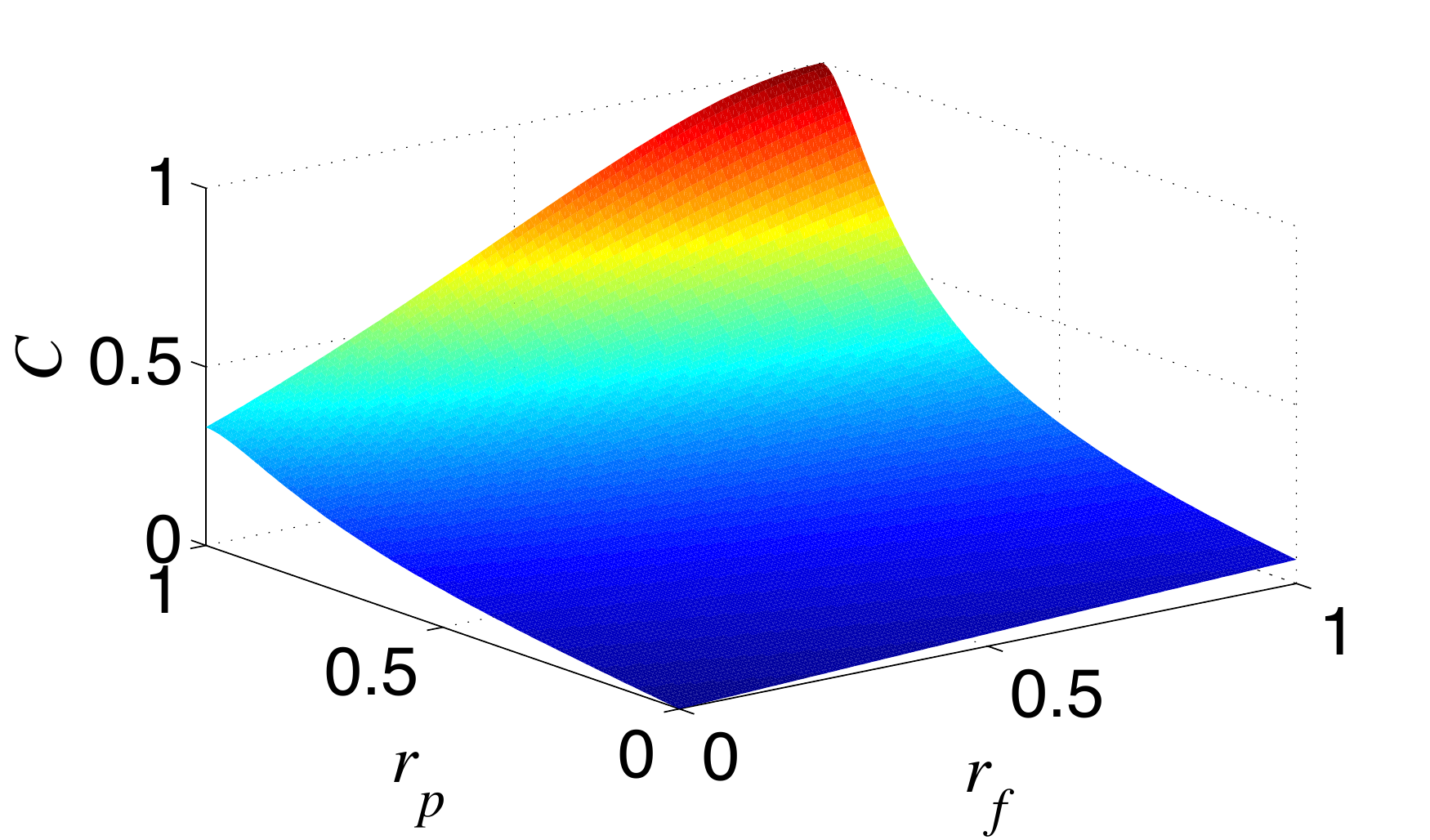}\figlab{conc_aniso_c}
   }
   \subfloat[$f_a=0.89$, $p_a=0.2$, $t_a=0.4$]{
     \includegraphics[width=0.49\columnwidth]{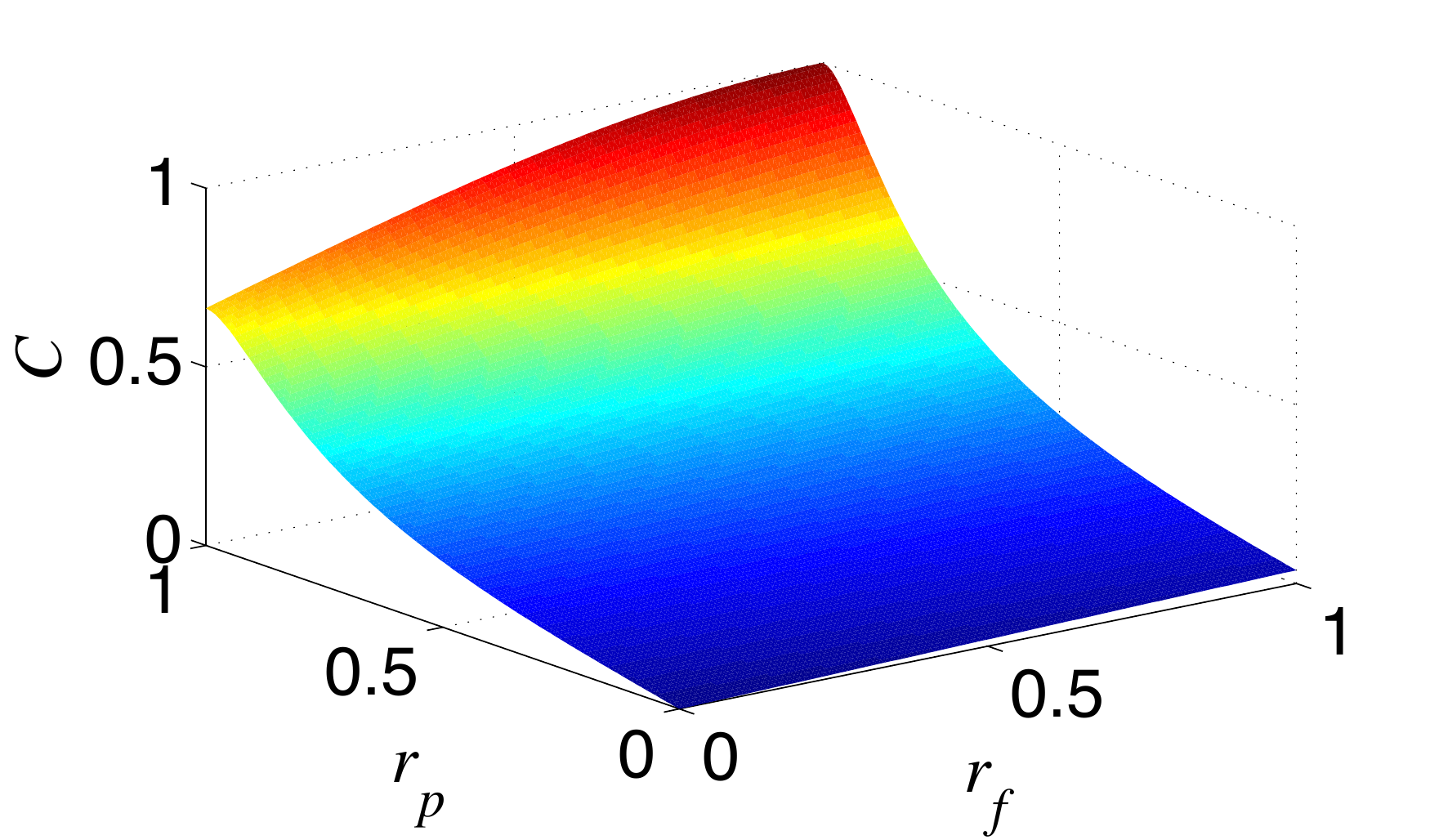}\figlab{conc_aniso_d}
     }\\
          \subfloat[$f_a=0.58$, $p_a=0.37$, $t_a=0.73$, $\Delta\phi_f=\Delta\phi_p=\Delta\phi_t=0$]{
        \includegraphics[width=0.49\columnwidth]{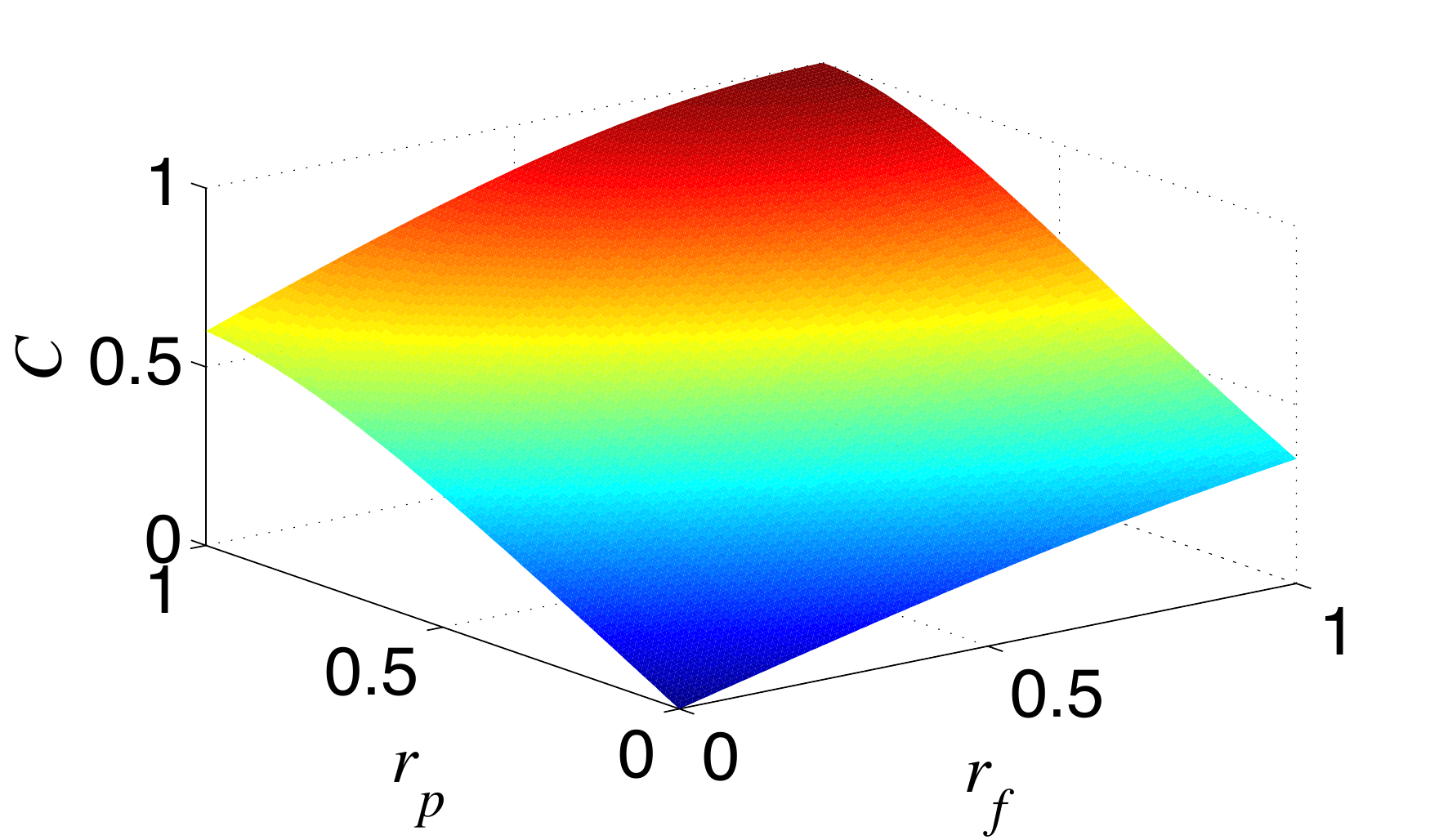}\figlab{conc_aniso_e}
   }
   \subfloat[$f_a=0.58$, $p_a=0.37$, $t_a=0.73$, $\Delta\phi_f=3\pi/2$, $\Delta\phi_p=\pi/5$, $\Delta\phi_t=4\pi/3$]{
     \includegraphics[width=0.49\columnwidth]{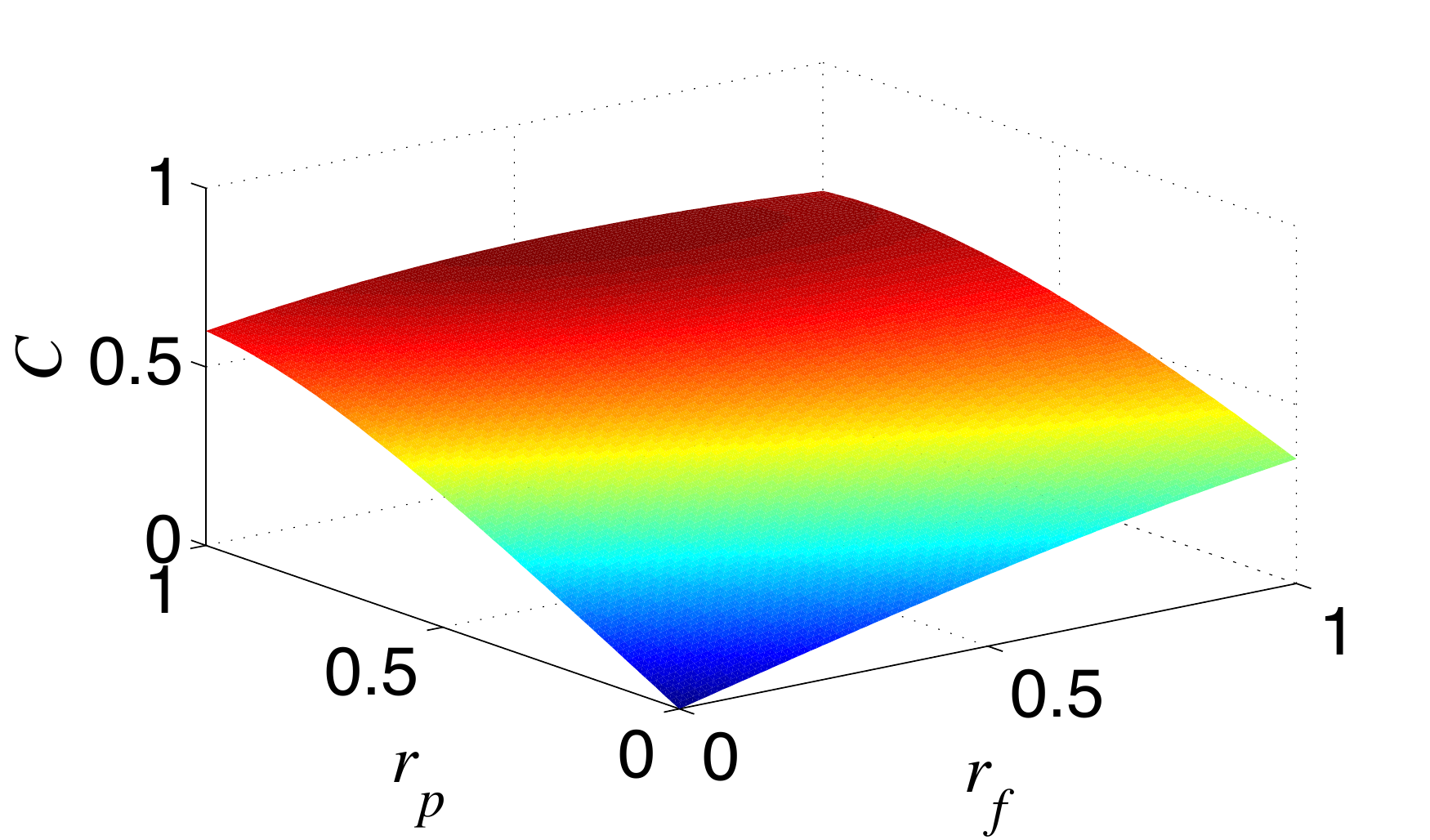}\figlab{conc_aniso_f}
     }

\caption{Maximal concurrence obtainable by tuning $V_g$, as a function of the ratios $r_f$ and $r_p$ between the different tunneling amplitudes in the two junctions. The difference between (e) and (f) illustrates the effect of allowing for differences in the phases added to the edge states during scattering in the two junctions.}
   \figlab{conc_aniso}
   \end{center}
   \end{figure}

 The five figures 4(a) - 4(e) are all plotted under the assumption that the differences in the phases acquired during tunneling in the junctions are zero. Most configurations of tunneling amplitudes allow for the $V_g$-controlled phases to compensate for finite phase differences to a large extent, but the concurrence is typically lowered somewhat. Let us denote by $\Delta\phi_f$, $\Delta\phi_p$, and $\Delta\phi_t$ the differences between the two junctions $a$ and $b$ in phases added to the edge states due to spin-flip tunneling, spin-preserving tunneling and transmission, respectively. In Fig.\ 4(f), these phase differences are chosen to destroy as much as possible of the entanglement for the case where the absolute values of the tunneling amplitudes are equal. It shows that the concurrence is lowered from $C=1$ to $C\approx 0.6$.

Having full control over the spin quantization axis of the entangled states is of course also highly desirable, in particular in a quantum information setting. As we have pointed out above, this requires control also over the spin-flip and spin-preserving amplitudes in the two tunneling junctions. In this context one should note that in the case of equal junctions $a$ and $b$, the concurrence in \Eqnref{cac} takes the form
\beq  \label{equal}
C=\frac{\left| P T\left(e^{2iKl_6}+F^2e^{-2iKl_6}\right)-F\left(P^2+T^2\right)\right|}{PT\left|1+Fe^{-2iKl_6}\right|^2+F \left(P+T\right)^2},
\eeq 

\noindent showing that for this particular case the amount of entanglement produced does not depend on the phases which the electrons acquire during tunneling. As expected, we also see from Eq. (\ref{equal}) that $C$ reaches unity for the cases with maximally entangled Bell pairs, choosing one of the possibilities $f=0$, $p=0$, or $t=0$.

It is also interesting to inquire into the role of the Aharonov-Bohm effect on the entanglement production. If the middle of the ring is threaded with a magnetic flux $\Phi$, the states will acquire an Aharonov-Bohm (A-B) phase $\phi_{\mathit A\mbox{-}B}$ \cite{ab1}. We choose the direction of the flux so that a full counterclockwise revolution produces the A-B phase $\phi_{\mathit A\mbox{-}B}=\Phi/\Phi_0$, where $\Phi_0=h/e$. In this case, the sign of the phase is dependent on the direction of propagation, and the reduced scattering matrix becomes
\beq
\tilde S\!  =\!\!  \begin{pmatrix} \! p_a e^{-iK(l_1+l_2)-i\phi_{12}} &\! \!  \! \! \!f_bt^*_a e^{-iK(l_2+l_4+l_6)+i\phi_{34}} \\ \!  f^*_at_b e^{-iK(l_1+l_3+l_6)-i\phi_{12}} & \! \! \! \! p^*_b e^{-iK(l_3+l_4)+i\phi_{34}} \\ \! -t_a e^{-iK(l_1+l_5)} &  \! \! \! \! \! \! \! \! f_bp^*_a e^{-iK(l_4+l_5+l_6)+i\phi_{\mathit A\mbox{-}B}} \\ \!  -f^*_ap_be^{-iK(l_1+l_6+l_7)-i\phi_{\mathit A\mbox{-}B}} & \! \! \! \! t^*_b e^{-iK(l_4+l_7)}\\
-f_a^*f_be^{-iK(l_1+l_4+l_6)} & 0\\
0 & -f_a^*f_b e^{-iK(l_1+l_4+l_6)}
\end{pmatrix}\! \!,
\eeq

\noindent where $\phi_{12}+\phi_{34}=\phi_{\mathit A\mbox{-}B}$, with $\phi_{12}$ and $\phi_{34}$ the parts of the total phase picked up from the upper and lower half of the ring, respectively. Using the notation of \Eqnref{app}, it follows that the outgoing state in the A-B case is, up to an unimportant global phase,
\begin{multline}
\ket{\Psi_{\textit out, A\mbox{-}B}}=N \Big(A_{\up\up} e^{i\phi_{\mathit A\mbox{-}B}}\ket{\up\up} + A_{\up\dn}\ket{\up\dn}\\
+ A_{\dn\up}e^{i\phi_{\mathit A\mbox{-}B}} \ket{\dn\up} + A_{\dn\dn} \ket{\dn\dn}\Big),
\end{multline}

\noindent and thus, according to \Eqnref{concurrence}, the concurrence is
\begin{multline}
C_{\mathit A\mbox{-}B}=2N^2\left| \left(A_{\up\dn}A_{\dn\up}-A_{\up\up}A_{\dn\dn}\right)e^{i\phi_{\mathit A\mbox{-}B}} \right|\\
=2N^2\left| A_{\up\dn}A_{\dn\up}-A_{\up\up}A_{\dn\dn} \right|,
\end{multline}

\noindent i.e.\ the same as without the A-B phase.

\section{Measurements}

 \begin{figure}
	\begin{center}
		\includegraphics[width=0.6 \columnwidth]{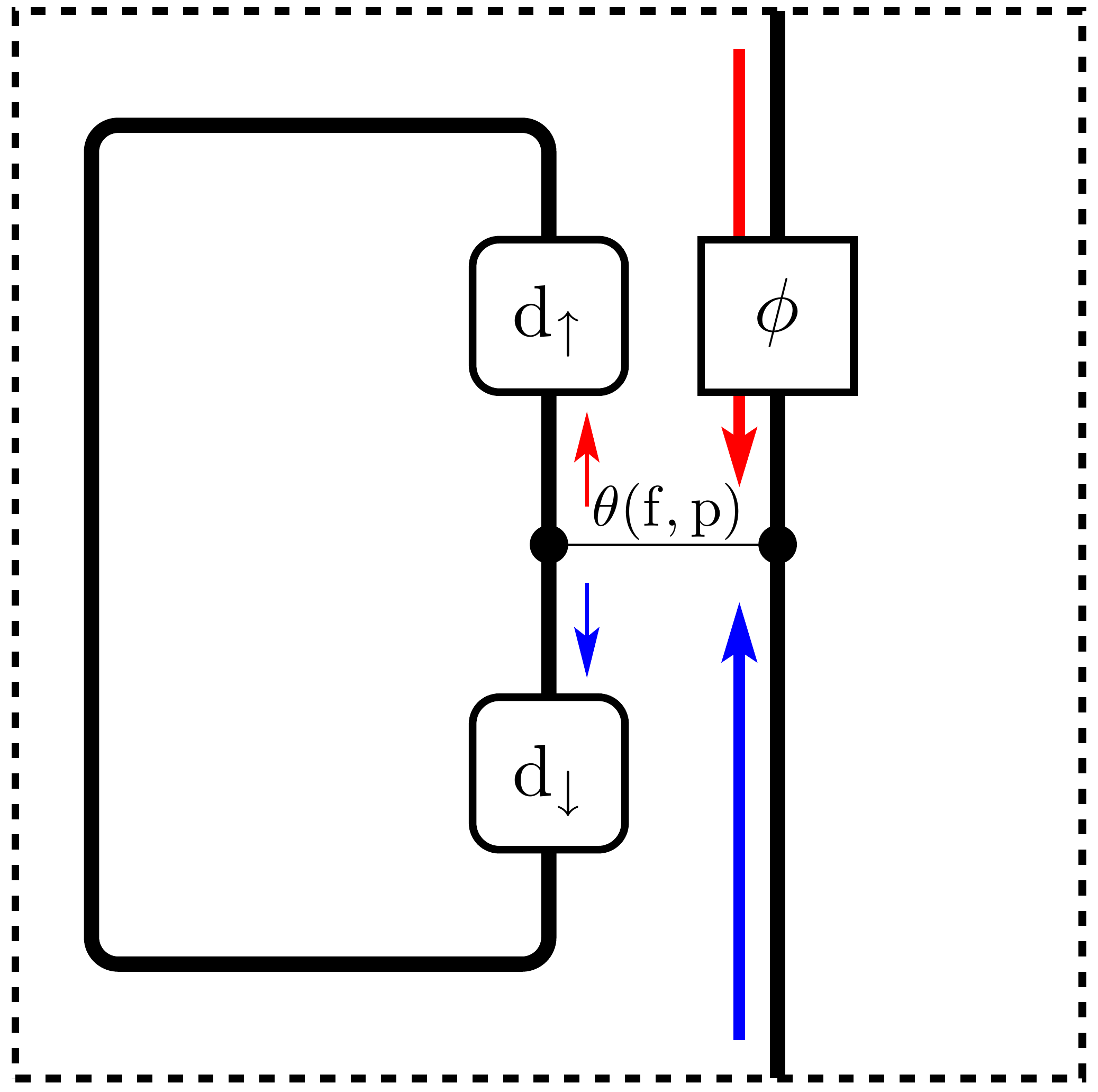}
		\caption{Sketch of a principle for measuring the entanglement produced in the system.  The angles $\theta$ and $\phi$ are effective rotation angles for the spin quantization axis, realized by using electric gates to control the spin-preserving and spin-flip tunneling and to add a phase to the states of spin up, respectively. The detectors $d_\up$ and $d_\dn$ detect electrons with spin up and spin down, respectively.}
		\figlab{measurement}
	\end{center} 
\end{figure}

\subsection{Bell test}

As discussed above, the spin entanglement in our device is produced through postselection of entangled electron pairs, discarding the pairs where either both electrons end up in the same detector, or where at least one of them returns back to the source. If a spin measurement is performed in the detectors, e.g.\ to assess the potential entanglement produced by the system, the very detection of the electrons gives us the possibility to disregard unwanted events. However, the measurement of the spins destroys the entanglement, so if we want to produce entangled pairs for quantum information processing, a noninvasive charge measurement that leaves the spins intact must first be performed in order to discard same-detector events \cite{Bose02, Zuelicke2005}. To achieve this, one may use a quantum point contact (QPC) capacitively coupled to the part of the leads where the electrons leave the entangler. The conductance through the QPC can be tuned to be extremely sensitive to nearby charges and this type of charge sensing is widely used in detecting single electrons on quantum dots \cite{fieldetal, schleseretal, vandersypenetal}. The precise setup of the charge measurement will of course depend on the setting in which the entangler will be used.

Measuring the nonlocal quantum correlations associated with the spin entanglement is a challenging task. It is here important to realize that the relative phases of the entangled spin states in Eq.\ (\ref{psiout}) imply that the spins in a generic state are rotated out of the $xz$ plane. When measuring the correlations between the states in the two detectors, we therefore need to be able to measure the spin along an arbitrary axis. This could be accomplished by having detectors where the spin-quantization axis is magnetically rotated in an arbitrary direction, parametrized by two angles $\theta$ and $\phi$. An example of this type of spin measurement is the use of a Zeeman-split quantum dot as a spin filter \cite{Recher2000, Hansonetal2004}.

A simpler solution in our case is to use detectors according to the sketch in Fig.\ \figref{measurement} (each of the detectors D$_1$ and D$_2$ in Fig. \figref{corbino} are now replaced by the setup within the dashed frame of Fig. \figref{measurement}), similar to the quantum state tomography setup of Ref. \cite{samuelssonQST}. Here, the role of the spin quantization axis is taken by a controllable tunnel junction with spin-flip and spin-preserving amplitudes $f$ and $p$, respectively. Since we want all electrons entering the detector to be measured, the amplitude for transmission should be $t=0$, and thus $|f|+|p|=1$. This means that we can introduce a single parameter $\theta$ to represent the control of the tunable tunneling amplitudes and define $\theta$ through the relations  $|f(\theta)|=\sin{\theta}$ and $|p(\theta)|=\cos{\theta}$. Tuning $\theta$ is then equivalent to rotating the spin-quantization axis in the $xz$ plane. Specifically, detecting an electron in detector $d_\up$ or $d_\dn$ becomes equivalent to measuring spin $\up$ or $\dn$ along the spin axis rotated by an angle $\theta$ around the original $y$ axis. A gate is placed along the path for one of the spins (here the $\uparrow$ spin), allowing us to impose an extra phase $e^{i\phi}$ to the states with spin up, equivalent to a rotation of the spin-quantization axis around the $z$ axis. The operators  $d_\up$ and $d_\dn$ of the detected states are then related to the operators $b_\up$ and $b_\dn$ of the outgoing states from the device through the relation
\beq
\begin{pmatrix}d_\up\\d_\dn\end{pmatrix} =\begin{pmatrix} p(\theta) e^{i\phi} & -f(\theta)\\ f(\theta) e^{i\phi} & p(\theta)\end{pmatrix} \begin{pmatrix} b_\up\\ b_\dn \end{pmatrix}.\eqnlab{spin3dmatrix}
\eeq

With these detectors, a Bell test can be carried out. Bell's inequality gives an upper bound for classical correlations between two states, and if it is violated we know that the states are quantum entangled. The correlations are expressed through the Bell parameter $B$, and the inequality is $B\leq2$. The version we will discuss here was proposed by Clauser, Horne, Shimony, and Holt (CHSH) \cite{chsh}. In this version the Bell parameter $B$ is calculated according to 
\begin{multline}
B=E(\theta_1,\phi_1,\theta_2,\phi_2)-E(\theta'_1,\phi'_1,\theta_2,\phi_2)\\
+E(\theta_1,\phi_1,\theta'_2,\phi'_2)-E(\theta'_1,\phi'_1,\theta'_2,\phi'_2),\eqnlab{b}
\end{multline}

\noindent where 
\beq
E(\theta_1,\phi_1,\theta_2,\phi_2)=P_{\up\dn}+P_{\dn\up}-P_{\up\up}-P_{\dn\dn}\eqnlab{e}
\eeq

\noindent and $P_{\sigma_1\sigma_2}=|A'_{\sigma_1\sigma_2}|^2$ is the probability to measure a state with spin $\sigma_1$ in D$_1$ and one with $\sigma_2$ in D$_2$ ($A'_{\sigma_1\sigma_2}$ being the amplitude). These probabilities depend on the chosen angles $\theta_i$, $\theta'_i$, $\phi_i$, and $\phi'_i, i=1,2$, in the corresponding detectors (cf. Fig. \figref{measurement}), and by varying them a maximal value  $B=B_{\mathit{max}}$ is obtained. For the pure states that we are considering $-$ having ensured that the two detected electrons were emitted with different spins $-$ the maximal value of $B$ is related to the concurrence through the relation $B_{\mathit{max}}=2\sqrt{1+C^2}$ 
\cite{bmaxref}. Using the notation in \Eqnref{app}, the amplitudes $A'_{\sigma_1\sigma_2}$ for the different outcomes are
\begin{multline}
A'_{\sigma_1\sigma_2}=A_{\up\up}\braket{\sigma_1\sigma_2}{\up\up}+A_{\up\dn}\braket{\sigma_1\sigma_2}{\up\dn}\\
+A_{\dn\up}\braket{\sigma_1\sigma_2}{\dn\up}+A_{\dn\dn}\braket{\sigma_1\sigma_2}{\dn\dn}.\eqnlab{bella}
\end{multline}

\noindent The matrix elements in \Eqnref{bella} should be read as $\braket{\sigma_1{\sigma_2}}{\lambda_1{\lambda_2}}=\left(\bra{d_{\sigma_1}}\otimes\bra{d_{\sigma_2}}\right)\left(\ket{b_{\lambda_1}}\otimes\ket{b_{\lambda_2}\right)}$, i.e.\ the amplitude for the state $\ket{b_{\lambda_1} b_{\lambda_2}}$ to be detected with spin $\sigma_1$ in D$_1$ and spin ${\sigma_2}$ in D$_2$. They can be calculated using \Eqnref{spin3dmatrix} so that, for example, $\braket{-+}{\up\dn}=f^*_1(\theta_1)f^*_2(\theta_2)e^{-i\phi_1}$, $\braket{--}{\dn\up}=-p^*_1(\theta_1)f^*_2(\theta_2)e^{-i\phi_2}$, and so on, where $+$ and $-$ denote spin up and down in the $d_\up$ and $d_\dn$ detectors, respectively.
With the source producing pairs of electrons of opposite spins well separated in time, the probabilities that enter the CHSH test described in Eqs.\ (\ref{eqn:b}) and (\ref{eqn:e}) are found by taking statistical averages of several events.

   \begin{figure}[ht!]
   \begin{center}
   \subfloat[$f=1/\sqrt{2}$, $p=t=1/(2\sqrt{2})$.\label{ell6}]{
   \includegraphics[width=0.9\columnwidth]{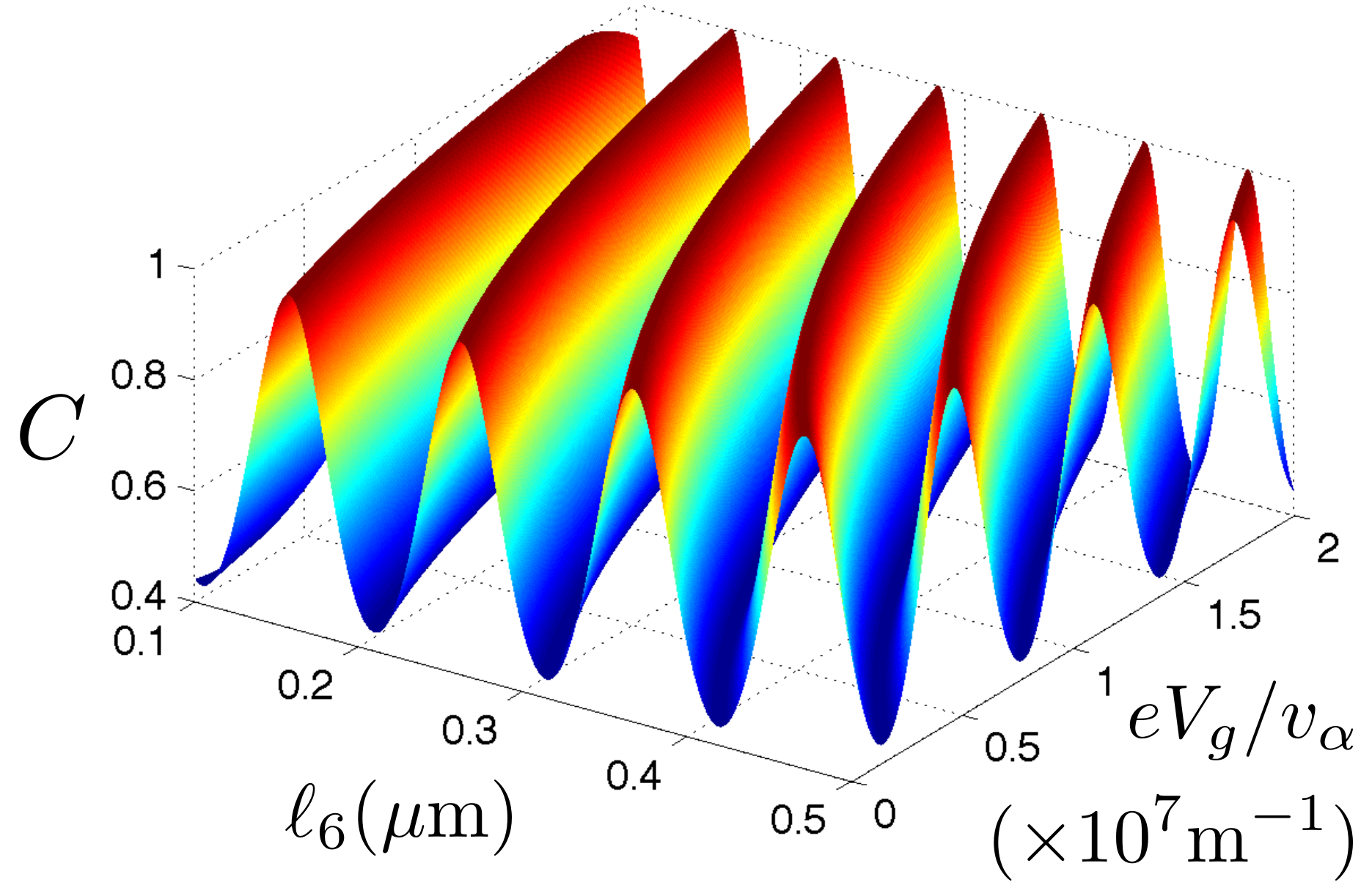}
   }
   \\
      \subfloat[$p=t=1/(2\sqrt{2})$, $l_6=400$ nm.\label{famp}]{
   \includegraphics[width=0.9\columnwidth]{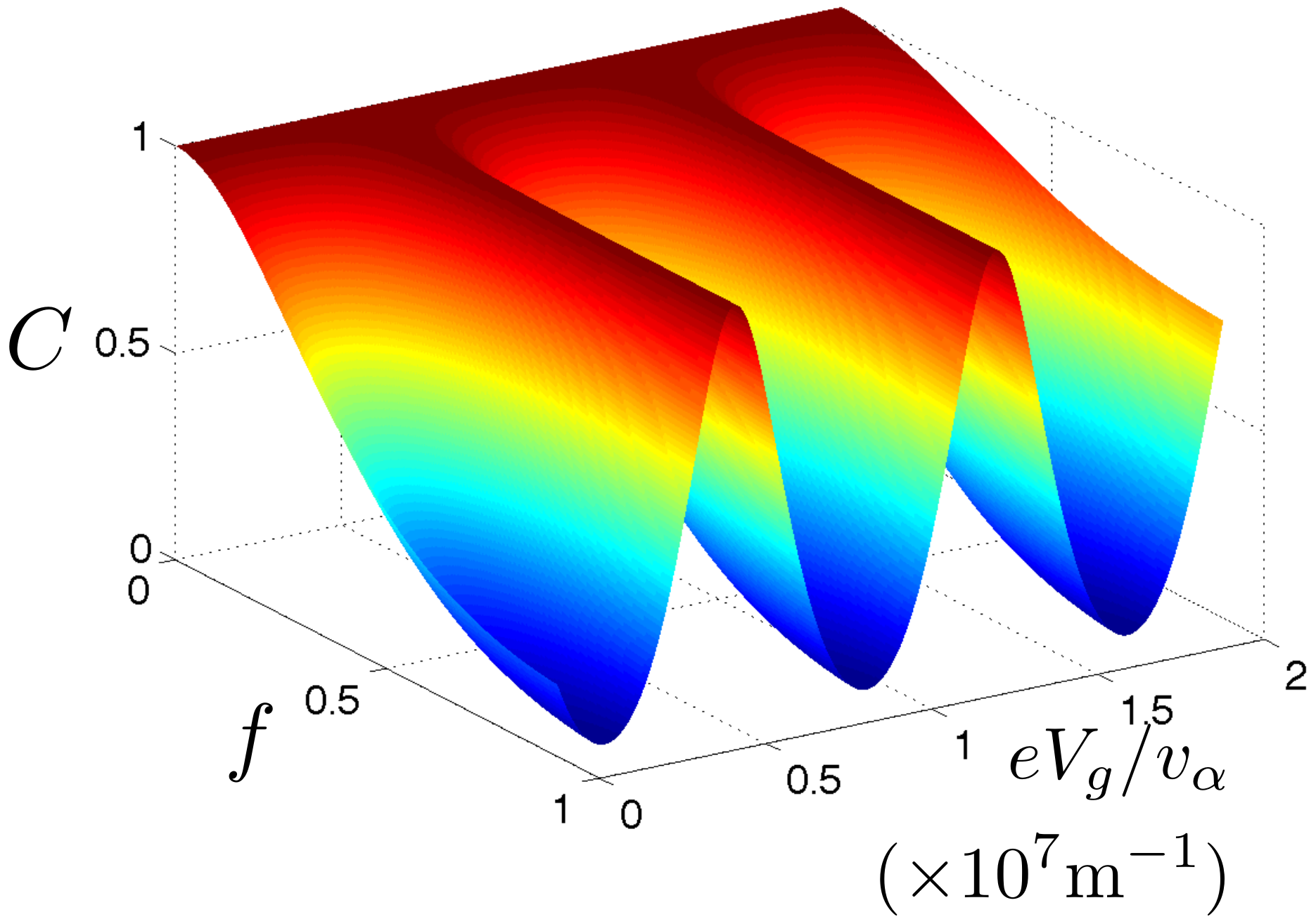}
   }\
   \caption{C as a function of $eV_g/v_\alpha$ and (a) $l_6$, (b) $|f|$, for the case of equal junctions. The oscillations in entanglement production due to the change in dynamical phase allow for using the effective gate potential $V_g$ and/or Rashba strength (via $v_\alpha$) to tune the concurrence to one with any set of tunneling amplitudes in the junctions, as long as they are equal in the two junctions $a$ and $b$.}
   \figlab{cplot}.
   \end{center}
   \end{figure}

\subsection{Experimental realization}

Most experiments on the QSH effect have been performed on HgTe/CdTe quantum wells \cite{qshexp}. In the aforementioned experiment where the Aharonov-Casher effect was measured in a ring structure in such a quantum well, a ring with an average radius of 1 $\mu$m was used \cite{acexp}. Conductance oscillations due to the phase change was observed with one period of oscillation for every change $\Delta V_g\approx 15$ mV of the gate voltage, estimated to correspond to tuning the Rashba SO splitting $\Delta_R$ between  $0<\Delta_R<50\ \mu \textrm{eV}$. Using an effective mass of $m^*=0.04\ m_e$, this is calculated to correspond to a tunable Rashba parameter in the interval $0<\alpha<4\times 10^{-12}$ eVm \cite{knigphase}. It should be noted that huge SO splittings of $\Delta_R=30$ meV have been observed in HgTe quantum wells with inverted band gaps \cite{guietal}. The width of the helical edge states are assumed to be approximately 40 nm \cite{qshexp}, so the thickness of the ring has to exceed, say, $\sim$ 100 nm to prevent unwanted overlaps across the edges.

With these numbers in mind, we can use \Eqnref{cac} to estimate the expected spin entanglement produced by our proposed device. Figure \figref{cplot} shows the concurrence $C$ in \Eqnref{cac} calculated as a function of  $eV_g/v_\alpha$ and (a) the length $l_6$, and (b) the absolute value of the spin-flip amplitude $|f|$. In both cases we have chosen $E=0$ in Eq.\ (\ref{energy}), and have also chosen identical junctions $a$ and $b$, with $f=f_a=f_b=1/\sqrt{2}$ and $p=p_a=p_b=t=t_a=t_b=1/(2\sqrt{2})$ in $(a)$, and $l_6=400$ nm in (b).  In (b), we have also chosen $p=t=1/(2\sqrt{2})$. The reason we plot $C$ as a function of $eV_g/v_{\alpha}$, rather than as a function of only $V_g$, is because of the complex dependence of $v_{\alpha}$ on $V_g$, being specific to the particular  design of the semiconductor heterostructure which supports the quantum well  \cite{MGJJ}. The oscillations in the entanglement production are shown clearly in Fig. \ref{ell6} and along the $eV_g/v_\alpha$ axis of Fig. \ref{famp}. Importantly, Fig. \ref{famp} shows that any spin-flip tunneling amplitude can produce maximum entanglement, given the right choice of effective gate potential and/or Rashba strength.
It is also interesting to note that the role of the relative sizes of spin-flip and spin-preserving amplitudes in the junctions when calculating the concurrence agrees qualitatively with the findings of Ref. \cite{inhoferbercioux}, where a related setup was considered in a topologically non-equivalent geometry. 

\section{Conclusion}

We considered a ring made from a quantum spin Hall insulator where a source injects pairs of electrons that are detected in two drains. Two beamsplitters in the ring with spin-flip and spin-preserving scattering paths result in spin-entangled portions of the wave functions where two electrons enter different detectors. Employing the process of postselection, spin-entanglement could be produced and then measured, e.g.\ via the violation of a Bell inequality, using spin-sensistive detectors. By exploiting the helical nature of the edge states, this detection may most effectively be carried out via a charge measurement. Most importantly, having a device with two equal beamsplitters allows for an electrical tuning of the output states between all four Bell states. This goes beyond earlier proposals based on chiral \cite{Samuelsson1, Samuelsson2, Beenakker} and helical \cite{inhoferbercioux} edge states. The calculated concurrence shows an oscillating behavior as a function of a dynamical phase which can be tuned via gate voltages. This dependence on the dynamical phase allows for electrical tuning of the amount of entanglement produced even in the case of an asymmetry between the two junctions. Postselection induces the entanglement in the first place, and the procedure could be used to create spin-entangled states useful for quantum information if the detection used for discarding unwanted events is spin-insensitive (i.e.\ only charge coincidences of the electrons are detected \cite{Bose02, Zuelicke2005}).
\\
We thank D. Bercioux, E. Eriksson, K. Le Hur, and P. Samuelsson for valuable discussions. HJ acknowledges CPHT at \'{E}cole Polytechnique for hospitality during the completion of this work. This research was supported by the Swedish Research Council under Grant No.\ 621-2011-3942, the DFG Grant No.\ RE 2978/1-1, and the EU-FP7 project SE2ND [271554].

% ----------------------------------------------------------------------------

\onecolumngrid
\appendix*
\subsection*{}

\section{The full $S$-matrix}

In order to make the time-reversal symmetry of the system manifest, we choose to write the scattering matrix in a basis where

\beq
\begin{pmatrix} b_{1\up}\\ b_{1\dn}\\ b_{2\up}\\ b_{2\dn}\\b_{S\up}\\ b_{S\dn}\end{pmatrix}= S \begin{pmatrix} a_{1\dn}\\ a_{1\up}\\ a_{2\dn}\\ a_{2\up}\\a_{S\dn}\\ a_{S\up}\end{pmatrix}.
\eeq

\noindent As defined in Sec. II.C, $a_{j \sigma} (b_{j \sigma}),$ j={\small S, D$_1$, D$_2$}; $\sigma = \up, \dn$, annihilates an electron in an ``incoming" (``outgoing") scattering state. By inspection of Fig. 1, and by imposing unitarity, the full $S$ matrix is thus obtained as

\begin{small}
\beq
S = \!\begin{pmatrix} \!0 & \!t^*_at_be^{-iK(l_2+l_3+l_6)} & \!-f_ae^{-iK(l_2+l_5)} & \!-t_a^*p_b e^{-iK(l_2+l_6+l_7)}  & \! t^*_af_b e^{-iK(l_2+l_4+l_6)}& \!p_a e^{-iK(l_1+l_2)} \\
\!t^*_at_be^{-iK(l_2+l_3+l_6)} &\! 0 & \!p^*_at_be^{-iK(l_3+l_5+l_6)} &\! f^*_be^{-iK(l_3+l_7)} & \!p^*_b e^{-iK(l_3+l_4)} &\!f^*_at_b e^{-iK(l_1+l_3+l_6)}\\
\!f_a e^{-iK(l_2+l_5)} & \!p^*_at_be^{-iK(l_3+l_5+l_6)} & \!0 & \!-p^*_ap_be^{-iK(l_5+l_6+l_7)} & \! p^*_af_b e^{-iK(l_4+l_5+l_6)} & \!-t_a e^{-iK(l_1+l_5)} \\
\!-t^*_ap_be^{-iK(l_2+l_6+l_7)} &\! -f^*_be^{-iK(l_3+l_7)} & \!-p^*_ap_be^{-iK(l_5+l_6+l_7)} & \!0  & \! t^*_b e^{-iK(l_4+l_7)}& \!-f^*_ap_be^{-iK(l_1+l_6+l_7)}\\
\!-t^*_af_be^{-iK(l_2+l_4+l_6)} &\! p^*_be^{-iK(l_3+l_4)} & \!-p^*_af_be^{-iK(l_4+l_5+l_6)} & \!t^*_be^{-iK(l_4+l_7)} & \!0 & \!-f^*_af_be^{-iK(l_1+l_4+l_6)}\\
\!p_ae^{-iK(l_1+l_2)} & -f^*_at_be^{-iK(l_1+l_3+l_6)} &\! -t_ae^{-iK(l_1+l_5)} &\! f^*_ap_be^{-iK(l_1+l_6+l_7)} &\! -f^*_af_be^{-iK(l_1+l_4+l_6)} &\! 0 \end{pmatrix}\!,\eqnlab{fullsmatrix}
\eeq
\end{small}

\noindent with all quantities entering the matrix elements defined in Sec. II. When extracting the reduced scattering matrix $\tilde S$ from $S$, it is convenient to pass to a new basis by switching columns $1\leftrightarrow 2$, $3 \leftrightarrow 4$ and $5\leftrightarrow 6$. Reading off from (A.2), this yields the expression for $\tilde S$ as in Eq. (\ref{tilde}).


\begin{thebibliography}{99}
\bibitem{footnote}
Note that the same arguments apply to the probability distributions in Eqs.\ (20) and (22), ensuring that the probabilities are indeed gauge invariant.
\end{thebibliography}
\end{document}